\documentclass[preprint2,numberedappendix]{emulateapj-rtx4}
\usepackage{graphicx,natbib,bm,color,amsmath,longtable,dashrule}
\graphicspath{{./fig/}{./png/}}

\newcommand{\EQ}{\begin{equation}}
\newcommand{\EN}{\end{equation}}
\newcommand{\EQA}{\begin{eqnarray}}
\newcommand{\ENA}{\end{eqnarray}}
\newcommand{\eq}[1]{(\ref{#1})}
\newcommand{\EEq}[1]{Equation~(\ref{#1})}
\newcommand{\Eq}[1]{Equation~(\ref{#1})}
\newcommand{\Eqs}[2]{Equations~(\ref{#1}) and~(\ref{#2})}

\newcommand{\Eqss}[2]{Equations~(\ref{#1})--(\ref{#2})}

\newcommand{\App}[1]{Appendix~\ref{#1}}
\newcommand{\Sec}[1]{Section~\ref{#1}}

\newcommand{\Fig}[1]{Figure~\ref{#1}}
\newcommand{\FFig}[1]{Figure~\ref{#1}}
\newcommand{\Figs}[2]{Figures~\ref{#1} and \ref{#2}}

\newcommand{\Tab}[1]{Table~\ref{#1}}

\newcommand{\bra}[1]{\langle #1\rangle}

{}
{}
{}
{}
{}

{}
{}
{}
{}
{}
{}
{}
{}
{}
{}
{}
{}
{}
{}
{}
{}
{}
{}
{}
{}
{}

{}
{}
{}

{}

{}
{}

\newcommand{\kk}{\bm{k}}

\newcommand{\xx}{\bm{x}}

\newcommand{\BB}{\bm{B}}

\newcommand{\uu}{\bm{u}}

\newcommand{\JJ}{\mbox{\boldmath $J$} {}}

\newcommand{\AAA}{\mbox{\boldmath $A$} {}}

\newcommand{\ee}{\mbox{\boldmath $e$} {}}

\newcommand{\nab}{\mbox{\boldmath $\nabla$} {}}

\newcommand{\SSSS}{\mbox{\boldmath ${\sf S}$} {}}
\newcommand{\hh}{\mbox{\boldmath ${\sf h}$} {}}
\newcommand{\TT}{\mbox{\boldmath ${\sf T}$} {}}
\newcommand{\TTT}{{\sf T}}
\newcommand{\PPP}{{\sf P}}
\newcommand{\eee}{{\sf e}}
\newcommand{\hhh}{{\sf h}}

\newcommand{\DD}{{\rm D} {}}

\newcommand{\dd}{{\rm d} {}}

\newcommand{\const}{{\rm const}  {}}

\def\Sp{\mbox{\rm Sp}}

\def\EEM{{\cal E}_{\rm M}}
\def\EEGW{{\cal E}_{\rm GW}}
\def\EGW{E_{\rm GW}}

\def\EM{E_{\rm M}}

\def\vmu{v_{\mu}}
\def\vlam{v_{\lambda}}

\def\xiM{\xi_{\rm M}}

\def\kone{k_1}
\def\kmu{k_\mu}

\def\kf{k_{\rm f}}

\def\EM{E_{\rm M}}
\def\kB{k_{\rm B}}

\def\kM{k_{\rm M}}

\def\Brms{B_{\rm rms}}

\def\half{{\textstyle{1\over2}}}

\newcommand{\G}{\,{\rm G}}
\newcommand{\Hz}{\,{\rm Hz}}

\newcommand{\uG}{\,\mu{\rm G}}
\newcommand{\K}{\,{\rm K}}

\newcommand{\cm}{\,{\rm cm}}

\newcommand{\Mpc}{\,{\rm Mpc}}

\newcommand{\GeV}{\,{\rm GeV}}

\newcommand{\yjhep}[3]{ #1, {JHEP,} {#2}, #3}
\newcommand{\yjcap}[3]{ #1, {JCAP,} {#2}, #3}

\newcommand{\yapj}[3]{ #1, {ApJ,} {#2}, #3}

\newcommand{\yapjl}[3]{ #1, {ApJ,} {#2}, #3}

\newcommand{\yana}[3]{ #1, {A\&A,} {#2}, #3}

\newcommand{\yanar}[3]{ #1, {A\&A Rev.,} {#2}, #3}

\newcommand{\ygafd}[3]{ #1, {Geophys.\ Astrophys.\ Fluid Dyn.,} {#2}, #3}

\newcommand{\yprl}[3]{ #1, {Phys.\ Rev.\ Lett.,} {#2}, #3}

\newcommand{\ymn}[3]{ #1, {MNRAS,} {#2}, #3}
\newcommand{\ynat}[3]{ #1, {Natur,} {#2}, #3}
\newcommand{\ysci}[3]{ #1, {Science,} {#2}, #3}

\newcommand{\ypr}[3]{ #1, {Phys.\ Rev.,} {#2}, #3}

\newcommand{\yprd}[3]{ #1, {PhRvD,} {#2}, #3}

\newcommand{\yjour}[4]{ #1, {#2}, {#3}, #4}

\newcommand{\pprr}[2]{ #1, {PhRvR}, in press, arXiv:#2}

\def\muz{{\mu}_{50}}
\def\muf{{\mu}_{5}}

\begin{document}

\title{Relic gravitational waves from the chiral magnetic effect}

\author{Axel~Brandenburg$^{1,2,3,4}$}
%\email{brandenb@nordita.org}

\author{Yutong~He$^{1,2}$}

\author{Tina Kahniashvili$^{3,4,5}$}
%\email{tinatin@andrew.cmu.edu}

\author{Matthias Rheinhardt$^{6}$}

\author{Jennifer~Schober$^{7}$}
%\email{jschober@nordita.org}

\affiliation{
$^1$Nordita, KTH Royal Institute of Technology and Stockholm University, Hannes Alfv\'ens v\"ag 12, SE-10691 Stockholm, Sweden \\
$^2$Department of Astronomy, AlbaNova University Center, Stockholm University, SE-10691 Stockholm, Sweden\\
$^3$McWilliams Center for Cosmology and Department of Physics, Carnegie Mellon University, 5000 Forbes Ave, Pittsburgh, PA 15213, USA\\
$^4$Faculty of Natural Sciences and Medicine, Ilia State University, 3-5 Cholokashvili Avenue, 0194 Tbilisi, Georgia\\
$^5$Department of Physics, Laurentian University, Ramsey Lake Road, Sudbury, ON P3E 2C, Canada\\
$^6$Department of Computer Science, Aalto University, PO Box 15400, FI-00076 Aalto, Finland\\
$^7$Laboratoire d'Astrophysique, EPFL, CH-1290 Sauverny, Switzerland
}

%\submitted{\today,~ $ $Revision: 1.263 $ $}
\submitted{Astrophys. J. 911, 110 (2021), \MakeLowercase{http://doi.org/10.3847/1538-4357/abe4d7}}
\date{Received 2021 January 18; accepted 2021 February 7; published 2021 April 22}

\begin{abstract}
Relic gravitational waves (GWs) can be produced by primordial magnetic
fields.
However, not much is known about the resulting GW amplitudes and their
dependence on the details of the generation mechanism.
Here we treat magnetic field generation through the chiral magnetic
effect (CME) as a generic mechanism and explore its dependence on the
speed of generation (the product of magnetic diffusivity
and characteristic wavenumber) and the speed characterizing the maximum
magnetic field strength expected from the CME.
When the latter exceeds the former (regime~I), the regime applicable to
the early universe, we obtain an inverse cascade with moderate GW energy
that scales with the third power of the magnetic energy.
When the generation speed exceeds the CME limit (regime~II), 
the GW energy continues to increase without a corresponding increase
of magnetic energy.
In the early kinematic phase, the GW energy spectrum (per linear
wavenumber interval) has opposite slopes in both regimes
and is characterized by an
inertial range spectrum in regime~I and a white noise spectrum in
regime~II.
The occurrence of these two slopes is shown to be a generic consequence
of a nearly monochromatic exponential growth of the magnetic field.
The resulting GW energy is found to be proportional to the fifth power
of the limiting CME speed and the first power of the generation speed.
\end{abstract}


\section{Introduction}
\label{Introduction}

The chiral magnetic effect (CME) describes an electric current along
a magnetic field carried by electrically charged chiral fermions
\citep{Vil80}.
This effect has been discussed as one of several possible mechanisms
for significantly amplifying primordial magnetic fields in the 
early universe \citep{BFR12,BFR15}.
It works as a dynamo effect that destabilizes the state of vanishing
magnetic field and causes an arbitrarily weak seed field to grow
exponentially for a limited time \citep{JS97}.
Excitation sets in when the fermion chiral asymmetry is large enough.
However, owing to the existence of a conservation law for the sum of
magnetic helicity and chiral asymmetry, the CME becomes continuously
depleted until nearly all the initial chiral asymmetry is turned into
magnetic helicity \citep{BFR12,BFR15}.
Thus, the initial chiral asymmetry determines the final value of the product of
the mean squared magnetic field $\Brms^2$ and the magnetic correlation
length $\xiM$, forming
a proxy for magnetic helicity in case of a fully helical field.
For realistic parameters describing our universe, 
$\Brms^2\xiM$ is expected to be of the order of or below
$(10^{-18}\G)^2\Mpc$ \citep{BSRKBFRK17}.
This value is below the lower limit of $\Brms^2\xiM>(10^{-16}\G)^2\Mpc$
that is inferred from the non-observations of GeV-energy halos around
TeV blazars \citep{Aharonian,NV10,TVN11}.
Yet the question can be raised, whether the resulting magnetic stress
could still be large enough to produce measurable gravitational waves
(GWs).

Another severe problem are the very small length scales 
associated with the CME.
An upper bound for the wavenumber associated with the chiral asymmetry
in comoving units is $k_*\equiv\kB T/\hbar c=12\cm^{-1}$, where
$\kB$ is the Boltzmann constant, $\hbar$ is the reduced Planck constant,
$c$ is the speed of light, and $T=2.7\K$ is the present day temperature.
Assuming a field strength of $1\uG$, 
the value of $k_*$ is compatible with the upper bound on
the magnetic helicity of $(10^{-18}\G)^2\Mpc$ \citep{BSRKBFRK17}.
This value of $k_*$ corresponds to very small length scales,
because the CME is a microphysical effect involving just $\hbar$, $c$,
and $\kB$ as relevant natural constants, but not Newton's constant
or the Planck mass; see also \cite{BKMRPTV17}.
The Hubble radius, by contrast, does involve Newton's constant and is
much bigger ($1.8\times10^{15}\cm$).
In units of the inverse Hubble radius, the characteristic scale of the 
CME corresponds to a
wavenumber of about $2\times10^{16}$; see Equation~(1) of \cite{KTBN13},
and is associated with a very high GW frequency of $4\times10^{11}\Hz$;
see Equation~(51) of \cite{KMK02}.  
On the other hand, at the time of the electroweak phase transition,
the Hubble scale corresponds to a frequency in the mHz range, which is
the range accessible to the Laser Interferometer Space Antenna.
Larger length scales have been argued to be possible by invoking
strongly out-of-equilibrium magnetic field generation during preheating
\citep{DiasGil08a,DiasGil08b}, or during inflation \citep{SSS19,OF21}.
In addition, the actual GW frequency could be several orders of magnitude
smaller owing to the inverse cascade associated with the CME.
By the time the magnetic field has reached its maximum, 
its typical length scale can therefore be significantly
larger than the scale at which the field was originally produced.
After that time, the magnetic length scales continue to increase as
the magnetic energy decreases.
However, \cite{RPMBKK20} found that the resulting GW energy is determined
just by the maximum field strength.
It is therefore unclear whether the late phase of magnetic decay
is still relevant to GW production.

Although the CME may not open a viable pathway for explaining the
primordial magnetic field, it has the advantage of providing a
self-consistent mechanism for explaining not just a certain field strength
and length scale, but also a certain time dependence of its generation,
independent of any extra assumptions.
Thus, it may serve as a proxy for other generation mechanisms.
It is then interesting to investigate GWs produced by the CME as
a mechanism that is likely to contain qualitatively valid aspects of
primordial magnetic field generation; see the recent work by \cite{ABPK19}
for analytic approaches addressing GW production from the CME at energies
much above the electroweak scale, or the approaches of \cite{SSS19}
and \cite{OF21} addressing GW production from helical magnetogenesis
during inflation.
These works give more optimistic prospects about the resulting magnetic
field generation than \cite{BSRKBFRK17}.
Therefore, in the present study
our aim is to understand the detailed relationship between the strengths
of magnetic field and GWs, as well as their typical time and length scales.

In the past, theoretical GW energy spectra have been calculated 
mostly using analytical approaches; see \cite{Deryagin} for an early
pioneering investigation and \cite{Caprini19} for a recent review.
Numerical approaches have recently been applied to GWs, driven by
acoustic turbulence from first order phase transitions \citep{HHRW15}.
A general uncertainty in simulating relic GWs from primordial turbulent
sources is due to our ignorance about suitable initial conditions
or generation mechanisms.
When a turbulent state is invoked as initial condition, the GW amplitude is
determined almost entirely by the fact that then the GW source, i.e.,
the turbulent stress, jumps instantaneously from zero to a finite value
\citep{RPBKKM20}.
By contrast, when driving turbulence gradually by applying some forcing
in the magnetohydrodynamic (MHD) equations, the resulting GW amplitude depends on
the details of how the turbulence develops and later declines;
see \cite{Kahn21} for a more systematic investigation.
These problems motivate our present study of GWs from the CME, too.

A number of interesting aspects of turbulence from the CME are already known.
In particular, depending on the relative rates of magnetic field
generation, on the one hand, and depletion of the CME, on the other,
different regimes of turbulence can be distinguished \citep{BSRKBFRK17}.
If the depletion is low,
the maximum magnetic field strength is high and
a turbulent spectrum with an inertial range emerges before the turbulence
starts to decay in a self-similar fashion.
If the depletion is high, on the other hand, no turbulent
inertial range develops.
How the resulting GW amplitude depends on the governing parameters of
the CME-driven field generation is unclear and illuminating this is the
main purpose of this paper.
Although the process is physically motivated, we choose parameters that
are motivated by our attempt to understand the relationship between
magnetic field generation and the resulting GWs
in any conceivable regime.
Our parameters are therefore not those relevant to the early
universe, nor are they necessarily physically realizable.
Nevertheless, the present work may prove to be important for guiding
our intuition about GW production from primordial turbulent sources.

\section{The model}

\subsection{Basic equations}
\label{BasicEquations}

The MHD equations for an ultrarelativistic quark-gluon plasma in a flat
expanding universe in the radiation-dominated era after the electroweak
phase transition can be written in terms of conformal time and comoving
coordinates such that the expansion no longer appears explicitly
\citep{BEO96,BKMRPTV17,DN13}, except for the GW equation; see below.
The bulk motions are assumed to be subrelativistic.

We quantify the chiral asymmetry through the
imbalance between the number densities $n_{\rm L}$
and $n_{\rm R}$ of left- and right-handed fermions, respectively, as
\EQ
\muf=24\,\alpha_{\rm em}\,(n_{\rm L}-n_{\rm R})\,
(\hbar c/k_{\rm B}T)^2,
\EN
employing the normalization used by \cite{Roga17}.
Here, $\alpha_{\rm em}$ is the fine structure constant.
The index 5 is commonly chosen in this context and reminiscent of
the fifth Dirac matrix $\gamma_5$, central in defining particle chirality.
We should point out that our $\muf$ has the unit of inverse length
and is related to the chiral chemical potential (with units of energy)
through an extra $\hbar c/4\alpha_{\rm em}$ factor; see \cite{Schober20}.  

We follow here the normalization of \cite{RPBKKM20,RPMBKK20}, where the
Heaviside-Lorentz system of units is used for the magnetic field and the
scale factor $a(t)$ is set to unity at the time $t_*$ of the electroweak
phase transition (denoted by an asterisk).
The Hubble parameter $H$ at $t_*$ is $H_*=t_*^{-1}$.
All quantities are made nondimensional by normalizing time by $t_*$,
velocities by the speed of light $c$, and the density $\rho$ by the
critical density $\rho_{\rm crit}$ for a flat universe.
Spatial coordinates are then normalized by the Hubble scale $c/H_*$.
Consequently, $\muf$ is normalized by $H_*/c$.
To obtain the comoving magnetic field in gauss, one has to multiply it
by $\sqrt{4\pi\rho_{\rm crit}}c$. 

The governing equations for the magnetic field $\BB$ and $\muf$
can then be written as \citep{Roga17,Schober18}
\begin{eqnarray}
{\partial\BB\over\partial t}&=&\nab\times[\uu\times\BB+\eta(\muf\BB-\JJ)],
\;\;\;\JJ=\nab\times\BB,\;
\label{dAdt}\\
{\DD\muf\over\DD t}&=&-\lambda\,\eta\left(\muf\BB-\JJ\right)\cdot\BB
+D_5\nabla^2\muf-\Gamma_{\rm\!f}\muf,
\label{dmudt}
\end{eqnarray}
where $\DD/\DD t\equiv\partial/\partial t+\uu\cdot\nab$ is the advective derivative,
$\eta$ is the magnetic diffusivity,
$\lambda$ characterizes the depletion of $\muf$ as the magnetic field increases, 
$D_5$ is a chiral diffusion coefficient,
and $\Gamma_{\rm\!f}$ is the flipping rate
\citep[see][for a recent calculation]{BCRS21}.
These equations have been derived under the assumption $\eta\to0$;
see \cite{Roga17} for details.
\cite{BSRKBFRK17} found that for $\kB T=100\GeV$
and if $\mu_{50}$ is produced thermally, relevant to the time
of the electroweak phase transition, $\Gamma_{\rm\!f}/\eta\muf^2\approx10^{-7}$,
that is, the time $1/\Gamma_{\rm\!f}$ is much longer than the e-folding time
of the fastest growing magnetic mode; see \Sec{basic}.
Hence, we put $\Gamma_{\rm\!f}=0$ from now on.
The plasma velocity $\uu$ and the density $\rho$ (which includes
the rest mass density) obey the momentum and energy equations
\begin{eqnarray}
{\DD\uu\over\DD t}&=&{2\over\rho}\nab\cdot\left(\rho\nu\SSSS\right)-{1\over4}\nab\ln\rho
+{\uu\over3}\left(\nab\cdot\uu+\uu\cdot\nab\ln\rho\right)
\nonumber \\
&-&{\uu\over\rho}\left[\uu\cdot(\JJ\times\BB)+\eta \JJ^2\right]
+{3\over4\rho}\JJ\times\BB,
\label{dudt} \\
{\partial\ln\rho\over\partial t}
&=&-\frac{4}{3}\left(\nab\cdot\uu+\uu\cdot\nab\ln\rho\right)
+{1\over\rho}\left[\uu\cdot(\JJ\times\BB)+\eta \JJ^2\right]\!,
\nonumber
\end{eqnarray}
where ${\sf S}_{ij}=(u_{i,j}+u_{j,i})/2 -\delta_{ij}\nab\cdot\uu/3$ are
the components of the rate-of-strain tensor with commas denoting partial
derivatives, $\nu$ is the kinematic viscosity, and the ultrarelativistic
equation of state $p = \rho/3$ has been employed.
In the following, we assume uniform $\nu$, $\eta$, and $D_5$ and vary
them such that $\nu=\eta=D_5$.

The GW equation in the radiation era for the scaled 
strain tensor $\hh$ with $\hhh_{ij} = a \hhh_{ij}^{\rm phys}$
is written in Fourier space as \citep{RPBKKM20,RPMBKK20}
\begin{equation}
\frac{\partial^2}{\partial t^2} \tilde{h}_{+/\times} (\kk, t) 
+k^2\tilde{h}_{+/\times} (\kk, t) = {6\over t} \tilde{T}_{+/\times}(\kk,t),
\label{GW4}
\end{equation}
where $\tilde{h}_{+/\times}=\eee_{ij}^{+/\times} (\PPP_{il}\PPP_{jm}-\half
\PPP_{ij}\PPP_{lm})\, \tilde{\hhh}_{lm}(\kk,t)$ are the Fourier-transformed
$+$ and $\times$ modes of  $\hh$, with $\eee^+_{ij} (\kk)\,\,=\,e_i^1 e_j^1 -
e_i^2 e_j^2$ and $\eee^\times_{ij} (\kk) \,=\,e_i^1 e_j^2 + e_i^2 e_j^1$
being the linear polarization basis,  $\ee^1$ and $\ee^2$ are unit
vectors perpendicular to $\kk$ and perpendicular to each other, and
$\PPP_{ij}(\kk) = \delta_{ij}-k_i k_j$ is the projection operator.
$\tilde{T}_{+/\times}$ are defined analogously and normalized by the
critical density.
The stress is composed of magnetic and kinetic contributions,
$\TTT_{ij} =\frac{4}{3}\gamma_{\rm Lor}^2\rho u_i u_j-B_i B_j+...$, where
$\gamma_{\rm Lor}=(1-\uu^2)^{-1/2}$ is the Lorentz factor, and
the ellipsis denotes terms proportional to $\delta_{ij}$, 
not contributing to $\tilde{T}_{+/\times}$.
Since we use the nonrelativistic equations, we put $\gamma_{\rm Lor}=1$, except
for one case shown in Appendix~\ref{ExtraTerm}, where $\gamma_{\rm Lor}\neq1$.
Our equations apply to the time after the electroweak phase transition
$t_\ast$, so our normalized time obeys $t\ge1$.
Furthermore, to compute the relic observable GW energy at the
present time, we have to multiply $\EEGW^{\rm sat}$ by
the square of the ratio of the Hubble parameters
and the fourth power of the ratio of scale factors
between the moment of the electroweak phase transition and today, which is
$1.64\times10^{-5}$; see \cite{RPBKKM20,RPMBKK20} for details.

As already alluded to above, the system of equations \eq{dAdt}, \eq{dmudt}
describing the CME must be regarded as partly phenomenological and subject to extensions
and modifications.
A purely helical magnetic field with wavenumber $k=\muf=\const$, for
example, can never decay if $\Gamma_{\rm\!f}=0$, and yet it would lead
to Ohmic heating.
However, those effects are not critical to the dynamics that we are
concerned with in this paper and will therefore be ignored.
Likewise, an extra $-\muf\nab\cdot\uu$ term on the right-hand side
of \Eq{dmudt} is necessary for a proper conservation equation.
However, this would not make a noticeable difference because
$\nab\cdot\uu$ is always small; see Appendix~\ref{ExtraTerm}
for a demonstration.
It should also be noted that, in comparison with earlier work,
this is the first time that the CME has been solved together with
Equations~\eq{dudt}, which contain additional 4/3 factors.
We refer to Appendix~A of \cite{BKMRPTV17} for the differences
to standard MHD.

\subsection{Basic phenomenology of the chiral magnetic effect}
\label{basic}
The CME introduces two important characteristic quantities into the
system: $\lambda$ and the initial value of $\muf$, 
$\muz = \muf(t=1)$, both assumed uniform. 
Different evolutionary scenarios can be envisaged depending on their values.
Following \cite{BSRKBFRK17}, we use the fact that $\lambda^{-1}$ has
the dimension of energy per unit length and $\muz$ has the dimension of
inverse length, and identify two characteristic velocities:
\EQ
\vlam=\muz/\lambda^{1/2},\quad\quad
\vmu=\muz\eta.
\EN
We recall that we have used here dimensionless quantities.
We can identify two regimes of interest:
\EQ
\eta \kone < \vmu <  \vlam \quad\mbox{(regime I)}, 
\label{regI}
\EN
\EQ
\eta \kone < \vlam <  \vmu \quad\mbox{(regime II)},
\label{regII}
\EN
where $k_1$ is the smallest wavenumber in the domain and
$\eta \kone < \vmu$ is necessary for magnetic field excitation.
The case $\vlam < \eta\kone$ is highly diffusive and was not considered.
In regime~I, if the ratio $\vlam/\vmu=\left[\eta\,\lambda^{1/2}\right]^{-1}$
is large, the $\lambda$ term is unimportant and $\muf$ will only change
slowly as the magnetic field grows.
Once the magnetic field exceeds a critical value of around
$\vmu$, it becomes turbulent; see \cite{BSRKBFRK17}.
In that paper, both $\vmu$ and $\vlam$ were assumed
to be less than the speed of sound, but this is not a physically
imposed constraint and will be relaxed in the present work.
\cite{BSRKBFRK17} also found that the crossover between the regimes
occurs when $\vlam/\vmu\approx8$.
Regarding the resulting GW production, however, we shall find
evidence for a crossover at $\vlam/\vmu\approx1$.
One should also remember that $\vmu$ and $\vlam$ do not correspond to
physically realizable speeds and are therefore not constrained to be
below unity.
Let us mention at this point that, using the calculation of
\cite{Arnold00} for the value of $\eta$ and
the expression $\lambda=3\hbar c\,(8\alpha_{\rm em}/\kB T)^2$ from \cite{Roga17},
\cite{BSRKBFRK17} estimated
that $\vmu\approx2\times10^{-5}$ and $\vlam\approx0.05$ for
$\muz=2\times10^{16}$.

If $\muz\ne0$, the CME determines primarily the magnetic helicity that
can subsequently be generated.
This is a direct consequence of the conservation law for the (weighted)
sum of mean magnetic helicity density and mean $\muf$, i.e.,
the {\em total mean chirality} \citep{Roga17},
\EQ
\half\lambda \, \bra{\AAA\cdot\BB}+\bra{\muf}=\const,
\label{totchirality}
\EN
where $\AAA$ with $\BB=\nab\times\AAA$ is the magnetic vector potential,
and the brackets denote averaging over a closed or periodic volume;
see \App{ExtraTerm} for a discussion of the accuracy of \Eq{totchirality}.
If the initial magnetic helicity is arbitrarily small, the constant in
\Eq{totchirality} can be set to $\muz$.
Neglecting the influence of the turbulent flow $\uu$ 
and inhomogeneities of $\muf$, the generated
magnetic field is fully helical (Beltrami), and its helicity
can be characterized by
its wavenumber $\kM$
and the mean magnetic energy density $\bra{\BB^2}/2$ through
$\bra{\AAA\cdot\BB}\approx\bra{\BB^2}/\kM$.
Therefore, once all the initial $\muf$ is used up, we have
\EQ
\bra{\BB^2}/\kM\approx2\muz/\lambda.
\label{B2xiM}
\EN
Interestingly, the value of $\eta$ does not enter this estimate.
It does, however, determine the initial growth rate
$\gamma(k)$ of the magnetic field, which adopts its maximum, 
$\gamma_0\equiv\eta\muz^2/4$, at the wavenumber $\kmu\equiv\muz/2$.
Using $\kM\approx\kmu$, we expect
\EQ
\bra{\BB^2}\lesssim\muz^2/\lambda\equiv\vlam^2,
\label{Bsat}
\EN
so large magnetic fields are expected for large values of $\muz$ and
small values of $\lambda$.
The fact that $\vlam$ characterizes the maximum magnetic field strength
justifies the name ``limiting CME speed''.
On the other hand, as one can express $\vmu$ by the maximum growth rate
and the corresponding wavenumber as $2\gamma_0/\kmu$, we may call it
``generation speed" in analogy to ``phase speed" for a wave.

\subsection{Magnetic energy spectrum from the CME}

To estimate the amount of magnetic energy production from the CME, we
adopt the semi-empirical model of \cite{BSRKBFRK17}, who proposed to
construct the magnetic energy spectrum such that it had the $k^{-2}$
slope that is characteristic of magnetically dominated turbulence,
with energy injection predominantly at the wavenumber $\kmu$.
For an intermediate time interval around the magnetic energy maximum,
they then proposed the following form for the magnetic energy spectrum
$\EM(k)$ (with normalization $\int \EM(k) dk = \bra{\BB^2}/2\equiv\EEM$)
as a function of wavenumber $k$ and the parameters $\eta$, $\muz$,
and $\lambda$ that govern the CME:
\EQ
E_{\rm M}(k)=C_5\,\muz^3\eta^2k^{-2}\quad  
\mbox{($k_\lambda\leq k\leq\kmu$)}
\label{Cmu}
\EN
where $C_5\approx16$ is a Kolmogorov-type constant, 
\EQ
k_\lambda=\sqrt{\lambda C_5/C_\lambda}\,\muz\eta
\approx4\muz\eta\lambda^{1/2}
\label{klambda}
\EN
is the wavenumber corresponding to the outer scale of the $k^{-2}$ subrange,
and $C_\lambda\approx1$ is another empirical constant \citep{BSRKBFRK17}.
Of course, \Eq{Cmu} can only hold if $k_\lambda\leq\kmu$.
In regime~I, $k_\lambda$ is the typical wavenumber of the magnetic field
when it has reached maximum strength.

A detailed sketch illustrating the different spectral subranges is
Figure~1 of \cite{BSRKBFRK17}, who also confirmed the form of \Eq{Cmu}
through simulations.
The present simulations also support the existence of the different
subranges.

\subsection{GW energy scaling}

The work of \cite{RPMBKK20} has shown that the GW energy
is not just proportional to the square of the magnetic energy, but
also proportional to the square of the dominating length scale (or
inverse wavenumber) of $\BB$.
For example, their Runs~ini2 and ini3 have the same magnetic energy,
but in ini3, the spectral peak was at a ten times smaller wavenumber,
corresponding to just ten turbulent eddies per Hubble horizon.
The resulting GW energy was then about a hundred times larger.
To leading order, the GW energy, normalized by the
critical energy of the universe, is given by
$\EEGW=\bra{\dot{h}_+^2+\dot{h}_\times^2}/6$;
see \cite{RPBKKM20} for details regarding the 1/6 factor
and additional correction terms.
\cite{RPMBKK20} studied different types of turbulence and confirmed
the quadratic relationship between the maximum magnetic energy,
$\EEM^{\max}$ and the saturation value of the GW energy,
$\EEGW^{\rm sat}$ in the form
\EQ
\EEGW^{\rm sat}\approx(q\EEM^{\max}/k_{\rm peak})^2,
\label{qdef}
\EN
where $k_{\rm peak}$ is the wavenumber of the peak of the spectrum
($k_{\rm peak}=600$ in most of their cases, and $60$ in the case where a
hundred times larger $\EEGW^{\rm sat}$ was found, suggesting an
inverse quadratic relationship), and $q$ is an empirical 
efficiency parameter that is about $0.9$ for their cases with a
turbulent initial MHD state (but no forcing),
$1.8$ in their simulations with forced MHD turbulence, and $11$ in their
simulations of forced
acoustic turbulence, where $\EEM^{\max}$ has
to be replaced by the maximum {\em kinetic} energy.
Larger values of $q$ correspond to more efficient conversion of
magnetic or kinetic energy into GW energy.
The reason why acoustic turbulence is more efficient is unclear,
but may be speculated to lie in its
more vigorous time dependence.

\subsection{Numerical aspects}

We solve \Eqss{dAdt}{GW4} using the {\sc Pencil Code} \citep{PC}, which
is a finite difference code that is third order in time and sixth order
in space, except that \Eq{GW4} is solved exactly between subsequent time
instants; see \citet{RPBKKM20} for details.
For most of our simulations, we use $512^3$ meshpoints, which turned out
to be sufficient for the present investigations.
The lowest wavenumber in our computational domain, $k_1$, is chosen to
be $100$ for many of our runs.
The side length of the cubical computational domain is then $2\pi/k_1$,
which is chosen to be large enough so that the governing dynamics
is well captured by the simulations, but small enough to resolve
the smallest length scales.
In many cases, we verified that the results are independent of
the choice of $k_1$.

Throughout his work, we present spectra of various quantities.
We denote this operation as $\Sp(\cdot)$, which is performed as
integration over concentric shells in wavenumber space.
For a scalar quantity $f$ it reads
$\Sp(f(\xx)) = k^2 \int |\tilde{f}(\kk)|^2 d \Omega_k$,
where $\Omega_k$ is the solid angle in $\kk$ space, 
while for the tensor $\hh$ we put
$\Sp(\hh) = \Sp(h_+) + \Sp(h_\times)$,
and likewise for $\dot{\hh}$ and $\TT$.
Thus, the GW energy spectrum is given by
$E_{\rm GW}(k)\equiv\Sp(\dot{\hh})/6$ and the magnetic one by 
$E_{\rm M}(k)\equiv[\Sp(B_x) + \Sp(B_y) + \Sp(B_z)]/2$.\footnote{
Let us note in this connection that one commonly denotes the GW energy
spectrum per {\em logarithmic} wavenumber interval by $\EEGW(k)$,
which is distinguished from $\EEGW$ by the argument $k$.
It is related to $E_{\rm GW}(k)$ through
$\EEGW(k)=kE_{\rm GW}(k)$.}

\section{Results}

We have performed a range of simulations where we vary $\eta$, $\lambda$,
and $\muz$, studying the influence of these parameters in turn.

\subsection{Comparison with earlier GW energy scaling}

\begin{figure}\begin{center}
\includegraphics[width=.95\columnwidth]{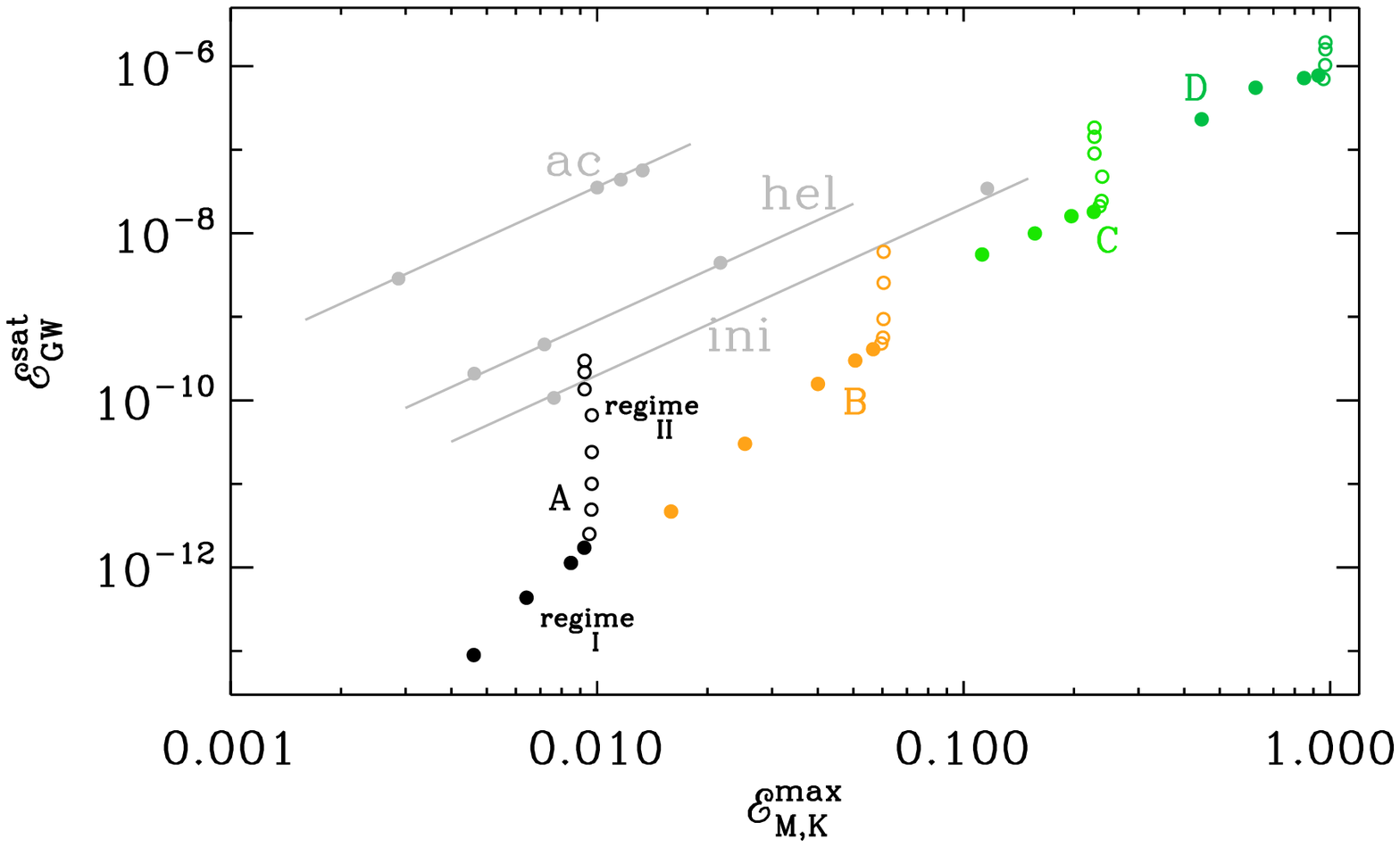}
\end{center}\caption{
$\EEGW^{\rm sat}$ versus $\EEM^{\max}$
for runs with $\muz=10^4$, grouped into four series with
$\lambda^{1/2}=5\times10^4$ (series~A), $2\times10^4$ (series~B),
$10^4$ (series~C), and $5\times10^3$ (series~D).
In each series we vary $\eta$.
Closed (open) circles refer to cases where $\vmu/\vlam<1$ ($>1$),
corresponding to regime~I (II).
For orientation, the data of \cite{RPMBKK20} are shown in gray; `ac' --
acoustic, `hel' -- helically forced MHD, `ini' -- 
turbulent initial MHD state (no forcing).
}\label{EEGW_vs_EEKM}\end{figure}

To put our new simulations into context, it is convenient
to compare our values of $\EEGW^{\rm sat}$ for given
$\EEM^{\max}$ with those obtained by \cite{RPMBKK20}.
We show in \Fig{EEGW_vs_EEKM} a plot similar to their Figure~7,
depicting our simulations with $\muz=10^4$, grouped into four series
with $\lambda^{1/2}$ in the range from $5\times10^4$ to $5\times10^3$.
In each of those series, we vary $\eta$.
The resulting values of $\EEM^{\max}$ and
$\EEGW^{\rm sat}$ are summarized in \Tab{Tsummary},
along with the four input parameters $\eta$, $\lambda^{1/2}$,
$\muz$, and $k_1$, as well as several derived quantities: $\eta k_1$,
$\vmu$, $\vlam$, $\eta\muz^2$, and $k_\lambda$ (provided $k_\lambda\leq\muz/2$).
In the last column, we also give according to \Eq{qdef}
\EQ
q=k_{\rm peak}\sqrt{\EEGW^{\rm sat}}/\EEM^{\max},
\label{qdef2}
\EN
where we estimate $k_{\rm peak}=k_\mu\min(1,\vmu/\vlam)$.
This means that $k_{\rm peak}=k_\mu$ when $\vmu > \vlam$ (regime~II)
and $k_{\rm peak}=k_\lambda/4$ when $\vmu < \vlam$ (regime~I);
see also \Eq{klambda}.

\begin{table*}\caption{
Summary of Runs from series A--G.
}\vspace{2pt}\centerline{\begin{tabular}{lccccccccccccc}
Run &$\eta$ & $\lambda^{1/2}$ &$\muz$ &$\eta k_1$& $\vmu$ & $\vlam$ & $\eta\muz^2$ & $k_1$ & $k_\lambda$ & $\EEM^{\max}$ & $\EEGW^{\rm sat}$ & $q$ \\[.5mm]
%Run   eta      lam1/2     mu       etak1      vmu       vlam       k1       klam    EEM       EEGW   q
\hline
A1&$1\times10^{-6}$&$5\times10^{4}$&$10^{4}$&$1\times10^{-4}$&$      0.01$&$       0.2$&$       100$&$       100$&$      2000$&$4.6\times10^{-3}$&$8.9\times10^{-14}$&$     0.032$\\
A2&$2\times10^{-6}$&$5\times10^{4}$&$10^{4}$&$2\times10^{-4}$&$      0.02$&$       0.2$&$       200$&$       100$&$      4000$&$6.4\times10^{-3}$&$4.3\times10^{-13}$&$      0.10$\\
A3&$5\times10^{-6}$&$5\times10^{4}$&$10^{4}$&$5\times10^{-4}$&$      0.05$&$       0.2$&$       500$&$       100$&$     (10000)$&$8.5\times10^{-3}$&$1.1\times10^{-12}$&$      0.31$\\
A4&$1\times10^{-5}$&$5\times10^{4}$&$10^{4}$&$1\times10^{-3}$&$       0.1$&$       0.2$&$      1000$&$       100$&---&$9.2\times10^{-3}$&$1.7\times10^{-12}$&$      0.71$\\[-2mm]
\multicolumn{14}{@{\hspace{0mm}}c@{\hspace{-2mm}}}{\hdashrule{167mm}{.5pt}{.7pt 2pt}}\\[-0.5mm]
A5&$2\times10^{-5}$&$5\times10^{4}$&$10^{4}$&$2\times10^{-3}$&$       0.2$&$       0.2$&$      2000$&$       100$&---&$9.5\times10^{-3}$&$2.5\times10^{-12}$&$       1.7$\\
A6&$5\times10^{-5}$&$5\times10^{4}$&$10^{4}$&$5\times10^{-3}$&$       0.5$&$       0.2$&$      5000$&$       100$&---&$9.6\times10^{-3}$&$4.9\times10^{-12}$&$       2.3$\\
A7&$1\times10^{-4}$&$5\times10^{4}$&$10^{4}$&$1\times10^{-2}$&$         1$&$       0.2$&$     10000$&$       100$&---&$9.7\times10^{-3}$&$1.0\times10^{-11}$&$       3.3$\\
A8&$2\times10^{-4}$&$5\times10^{4}$&$10^{4}$&$2\times10^{-2}$&$         2$&$       0.2$&$     20000$&$       100$&---&$9.7\times10^{-3}$&$2.4\times10^{-11}$&$       5.1$\\
A9&$5\times10^{-4}$&$5\times10^{4}$&$10^{4}$&$5\times10^{-2}$&$         5$&$       0.2$&$     50000$&$       100$&---&$9.7\times10^{-3}$&$6.6\times10^{-11}$&$       8.4$\\
A10&$1\times10^{-3}$&$5\times10^{4}$&$10^{4}$&$5\times10^{-2}$&$        10$&$       0.2$&$1\times10^{5}$&$        50$&---&$9.2\times10^{-3}$&$1.4\times10^{-10}$&$        12$\\
A11&$2\times10^{-3}$&$5\times10^{4}$&$10^{4}$&$1\times10^{-1}$&$        20$&$       0.2$&$2\times10^{5}$&$        50$&---&$9.2\times10^{-3}$&$2.2\times10^{-10}$&$        15$\\
A12&$5\times10^{-3}$&$5\times10^{4}$&$10^{4}$&$2\times10^{-1}$&$        50$&$       0.2$&$5\times10^{5}$&$        50$&---&$9.2\times10^{-3}$&$3.0\times10^{-10}$&$        18$\\
\hline
B1&$1\times10^{-6}$&$2\times10^{4}$&$10^{4}$&$1\times10^{-4}$&$      0.01$&$       0.5$&$       100$&$       100$&$       800$&$1.6\times10^{-2}$&$4.7\times10^{-12}$&$     0.027$\\
B2&$2\times10^{-6}$&$2\times10^{4}$&$10^{4}$&$2\times10^{-4}$&$      0.02$&$       0.5$&$       200$&$       100$&$      1600$&$2.5\times10^{-2}$&$3.0\times10^{-11}$&$     0.087$\\
B3&$5\times10^{-6}$&$2\times10^{4}$&$10^{4}$&$5\times10^{-4}$&$      0.05$&$       0.5$&$       500$&$       100$&$      4000$&$4.0\times10^{-2}$&$1.6\times10^{-10}$&$      0.31$\\
B4&$1\times10^{-5}$&$2\times10^{4}$&$10^{4}$&$1\times10^{-3}$&$       0.1$&$       0.5$&$      1000$&$       100$&$      (8000)$&$5.1\times10^{-2}$&$3.0\times10^{-10}$&$      0.68$\\
B5&$2\times10^{-5}$&$2\times10^{4}$&$10^{4}$&$2\times10^{-3}$&$       0.2$&$       0.5$&$      2000$&$       100$&---&$5.7\times10^{-2}$&$4.1\times10^{-10}$&$       1.4$\\[-2mm]
\multicolumn{14}{@{\hspace{0mm}}c@{\hspace{-2mm}}}{\hdashrule{167mm}{.5pt}{.7pt 2pt}}\\[-0.5mm]
B6&$5\times10^{-5}$&$2\times10^{4}$&$10^{4}$&$5\times10^{-3}$&$       0.5$&$       0.5$&$      5000$&$       100$&---&$6.0\times10^{-2}$&$4.8\times10^{-10}$&$       3.7$\\
B7&$1\times10^{-4}$&$2\times10^{4}$&$10^{4}$&$1\times10^{-2}$&$         1$&$       0.5$&$     10000$&$       100$&---&$6.0\times10^{-2}$&$5.6\times10^{-10}$&$       3.9$\\
B8&$2\times10^{-4}$&$2\times10^{4}$&$10^{4}$&$2\times10^{-2}$&$         2$&$       0.5$&$     20000$&$       100$&---&$6.0\times10^{-2}$&$9.4\times10^{-10}$&$       5.1$\\
B9&$5\times10^{-4}$&$2\times10^{4}$&$10^{4}$&$5\times10^{-2}$&$         5$&$       0.5$&$     50000$&$       100$&---&$6.0\times10^{-2}$&$2.6\times10^{-9}$&$       8.4$\\
B10&$1\times10^{-3}$&$2\times10^{4}$&$10^{4}$&$1\times10^{-1}$&$        10$&$       0.5$&$1\times10^{5}$&$       100$&---&$6.0\times10^{-2}$&$6.0\times10^{-9}$&$        12$\\
\hline
C1&$5\times10^{-6}$&$10^{4}$&$10^{4}$&$5\times10^{-4}$&$      0.05$&$         1$&$       500$&$       100$&$      2000$&$1.1\times10^{-1}$&$5.6\times10^{-9}$&$      0.33$\\
C2&$1\times10^{-5}$&$10^{4}$&$10^{4}$&$1\times10^{-3}$&$       0.1$&$         1$&$      1000$&$       100$&$      4000$&$1.6\times10^{-1}$&$9.9\times10^{-9}$&$      0.64$\\
C3&$2\times10^{-5}$&$10^{4}$&$10^{4}$&$2\times10^{-3}$&$       0.2$&$         1$&$      2000$&$       100$&$      (8000)$&$2.0\times10^{-1}$&$1.6\times10^{-8}$&$       1.3$\\
C4&$5\times10^{-5}$&$10^{4}$&$10^{4}$&$5\times10^{-3}$&$       0.5$&$         1$&$      5000$&$       100$&---&$2.3\times10^{-1}$&$1.8\times10^{-8}$&$       3.0$\\[-2mm]
\multicolumn{14}{@{\hspace{0mm}}c@{\hspace{-2mm}}}{\hdashrule{167mm}{.5pt}{.7pt 2pt}}\\[-0.5mm]
C5&$1\times10^{-4}$&$10^{4}$&$10^{4}$&$1\times10^{-2}$&$         1$&$         1$&$     10000$&$       100$&---&$2.3\times10^{-1}$&$2.1\times10^{-8}$&$       6.2$\\
C6&$2\times10^{-4}$&$10^{4}$&$10^{4}$&$2\times10^{-2}$&$         2$&$         1$&$     20000$&$       100$&---&$2.4\times10^{-1}$&$2.4\times10^{-8}$&$       6.6$\\
C7&$5\times10^{-4}$&$10^{4}$&$10^{4}$&$5\times10^{-2}$&$         5$&$         1$&$     50000$&$       100$&---&$2.4\times10^{-1}$&$4.8\times10^{-8}$&$       9.1$\\
C8&$1\times10^{-3}$&$10^{4}$&$10^{4}$&$5\times10^{-2}$&$        10$&$         1$&$1\times10^{5}$&$        50$&---&$2.3\times10^{-1}$&$9.0\times10^{-8}$&$        13$\\
C9&$2\times10^{-3}$&$10^{4}$&$10^{4}$&$1\times10^{-1}$&$        20$&$         1$&$2\times10^{5}$&$        50$&---&$2.3\times10^{-1}$&$1.4\times10^{-7}$&$        16$\\
C10&$5\times10^{-3}$&$10^{4}$&$10^{4}$&$2\times10^{-1}$&$        50$&$         1$&$5\times10^{5}$&$        50$&---&$2.3\times10^{-1}$&$1.8\times10^{-7}$&$        18$\\
\hline
D1&$1\times10^{-5}$&$5\times10^{3}$&$10^{4}$&$2\times10^{-3}$&$       0.1$&$         2$&$      1000$&$       200$&$      2000$&$4.5\times10^{-1}$&$2.3\times10^{-7}$&$      0.54$\\
D2&$2\times10^{-5}$&$5\times10^{3}$&$10^{4}$&$2\times10^{-3}$&$       0.2$&$         2$&$      2000$&$       100$&$      4000$&$6.3\times10^{-1}$&$5.5\times10^{-7}$&$       1.2$\\
D3&$5\times10^{-5}$&$5\times10^{3}$&$10^{4}$&$5\times10^{-3}$&$       0.5$&$         2$&$      5000$&$       100$&$     (10000)$&$8.5\times10^{-1}$&$7.2\times10^{-7}$&$       2.5$\\
D4&$1\times10^{-4}$&$5\times10^{3}$&$10^{4}$&$1\times10^{-2}$&$         1$&$         2$&$     10000$&$       100$&---&$9.3\times10^{-1}$&$7.7\times10^{-7}$&$       4.7$\\[-2mm]
\multicolumn{14}{@{\hspace{0mm}}c@{\hspace{-2mm}}}{\hdashrule{167mm}{.5pt}{.7pt 2pt}}\\[-0.5mm]
D5&$2\times10^{-4}$&$5\times10^{3}$&$10^{4}$&$4\times10^{-2}$&$         2$&$         2$&$     20000$&$       200$&---&$9.6\times10^{-1}$&$7.0\times10^{-7}$&$       8.7$\\
D6&$5\times10^{-4}$&$5\times10^{3}$&$10^{4}$&$1\times10^{-1}$&$         5$&$         2$&$     50000$&$       200$&---&$9.7\times10^{-1}$&$1.0\times10^{-6}$&$        10$\\
D7&$1\times10^{-3}$&$5\times10^{3}$&$10^{4}$&$2\times10^{-1}$&$        10$&$         2$&$1\times10^{5}$&$       200$&---&$9.7\times10^{-1}$&$1.6\times10^{-6}$&$        12$\\
D8&$2\times10^{-3}$&$5\times10^{3}$&$10^{4}$&$4\times10^{-1}$&$        20$&$         2$&$2\times10^{5}$&$       200$&---&$9.7\times10^{-1}$&$1.9\times10^{-6}$&$        14$\\
\hline
E1&$5\times10^{-6}$&$10^{4}$&$2\times10^{4}$&$1\times10^{-3}$&$       0.1$&$         2$&$      2000$&$       200$&$      4000$&$4.4\times10^{-1}$&$8.6\times10^{-8}$&$      0.67$\\
E2&$1\times10^{-5}$&$10^{4}$&$2\times10^{4}$&$2\times10^{-3}$&$       0.2$&$         2$&$      4000$&$       200$&$      8000$&$6.3\times10^{-1}$&$1.5\times10^{-7}$&$       1.2$\\
E3&$2\times10^{-5}$&$10^{4}$&$2\times10^{4}$&$4\times10^{-3}$&$       0.4$&$         2$&$      8000$&$       200$&$     (16000)$&$8.1\times10^{-1}$&$1.9\times10^{-7}$&$       2.2$\\
E4&$5\times10^{-5}$&$10^{4}$&$2\times10^{4}$&$1\times10^{-2}$&$         1$&$         2$&$     20000$&$       200$&---&$9.3\times10^{-1}$&$2.0\times10^{-7}$&$       4.9$\\[-2mm]
\multicolumn{14}{@{\hspace{0mm}}c@{\hspace{-2mm}}}{\hdashrule{167mm}{.5pt}{.7pt 2pt}}\\[-0.5mm]
E5&$1\times10^{-4}$&$10^{4}$&$2\times10^{4}$&$2\times10^{-2}$&$         2$&$         2$&$     40000$&$       200$&---&$9.6\times10^{-1}$&$2.1\times10^{-7}$&$       9.6$\\
E6&$2\times10^{-4}$&$10^{4}$&$2\times10^{4}$&$4\times10^{-2}$&$         4$&$         2$&$     80000$&$       200$&---&$9.7\times10^{-1}$&$2.6\times10^{-7}$&$        10$\\
E7&$5\times10^{-4}$&$10^{4}$&$2\times10^{4}$&$1\times10^{-1}$&$        10$&$         2$&$2\times10^{5}$&$       200$&---&$9.7\times10^{-1}$&$4.6\times10^{-7}$&$        13$\\
E8&$1\times10^{-3}$&$10^{4}$&$2\times10^{4}$&$2\times10^{-1}$&$        20$&$         2$&$4\times10^{5}$&$       200$&---&$9.7\times10^{-1}$&$6.2\times10^{-7}$&$        16$\\
\hline
F1&$5\times10^{-6}$&$10^{3}$&$2\times10^{3}$&$2\times10^{-4}$&$      0.01$&$         2$&$        20$&$        50$&$        40$&$9.4\times10^{-2}$&$7.2\times10^{-11}$&$   0.00091$\\
F2&$1\times10^{-5}$&$10^{3}$&$2\times10^{3}$&$5\times10^{-4}$&$      0.02$&$         2$&$        40$&$        50$&$        80$&$1.4\times10^{-1}$&$6.7\times10^{-9}$&$     0.012$\\
F3&$2\times10^{-5}$&$10^{3}$&$2\times10^{3}$&$1\times10^{-3}$&$      0.04$&$         2$&$        80$&$        50$&$       160$&$2.5\times10^{-1}$&$1.7\times10^{-7}$&$     0.067$\\
F4&$5\times10^{-5}$&$10^{3}$&$2\times10^{3}$&$1\times10^{-3}$&$       0.1$&$         2$&$       200$&$        25$&$       400$&$4.5\times10^{-1}$&$3.3\times10^{-6}$&$      0.41$\\
F5&$1\times10^{-4}$&$10^{3}$&$2\times10^{3}$&$2\times10^{-3}$&$       0.2$&$         2$&$       400$&$        25$&$       800$&$6.3\times10^{-1}$&$8.2\times10^{-6}$&$      0.91$\\
F6&$2\times10^{-4}$&$10^{3}$&$2\times10^{3}$&$5\times10^{-3}$&$       0.4$&$         2$&$       800$&$        25$&$      (1600)$&$8.1\times10^{-1}$&$1.3\times10^{-5}$&$       1.8$\\
F7&$5\times10^{-4}$&$10^{3}$&$2\times10^{3}$&$1\times10^{-2}$&$         1$&$         2$&$      2000$&$        25$&---&$9.3\times10^{-1}$&$1.6\times10^{-5}$&$       4.3$\\[-2mm]
\multicolumn{14}{@{\hspace{0mm}}c@{\hspace{-2mm}}}{\hdashrule{167mm}{.5pt}{.7pt 2pt}}\\[-0.5mm]
F8&$1\times10^{-3}$&$10^{3}$&$2\times10^{3}$&$2\times10^{-2}$&$         2$&$         2$&$      4000$&$        25$&---&$9.6\times10^{-1}$&$1.8\times10^{-5}$&$       8.9$\\
\hline
G1&$1\times10^{-5}$&$5\times10^{2}$&$10^{3}$&$2\times10^{-4}$&$      0.01$&$         2$&$        10$&$        25$&$        20$&$9.8\times10^{-2}$&$1.1\times10^{-10}$&$   0.00054$\\
G2&$2\times10^{-5}$&$5\times10^{2}$&$10^{3}$&$5\times10^{-4}$&$      0.02$&$         2$&$        20$&$        25$&$        40$&$1.5\times10^{-1}$&$7.5\times10^{-9}$&$    0.0060$\\
G3&$5\times10^{-5}$&$5\times10^{2}$&$10^{3}$&$1\times10^{-3}$&$      0.05$&$         2$&$        50$&$        25$&$       100$&$2.9\times10^{-1}$&$6.0\times10^{-7}$&$     0.066$\\
G4&$1\times10^{-4}$&$5\times10^{2}$&$10^{3}$&$2\times10^{-3}$&$       0.1$&$         2$&$       100$&$        25$&$       200$&$4.5\times10^{-1}$&$4.8\times10^{-6}$&$      0.24$\\
G5&$2\times10^{-4}$&$5\times10^{2}$&$10^{3}$&$5\times10^{-3}$&$       0.2$&$         2$&$       200$&$        25$&$       400$&$6.3\times10^{-1}$&$1.6\times10^{-5}$&$      0.63$\\
G6&$5\times10^{-4}$&$5\times10^{2}$&$10^{3}$&$1\times10^{-2}$&$       0.5$&$         2$&$       500$&$        25$&$      (1000)$&$8.6\times10^{-1}$&$3.9\times10^{-5}$&$       1.8$\\
G7&$1\times10^{-3}$&$5\times10^{2}$&$10^{3}$&$1\times10^{-2}$&$         1$&$         2$&$      1000$&$        10$&---&$9.3\times10^{-1}$&$5.4\times10^{-5}$&$       4.0$\\[-2mm]
\multicolumn{14}{@{\hspace{0mm}}c@{\hspace{-2mm}}}{\hdashrule{167mm}{.5pt}{.7pt 2pt}}\\[-0.5mm]
G8&$2\times10^{-3}$&$5\times10^{2}$&$10^{3}$&$2\times10^{-2}$&$         2$&$         2$&$      2000$&$        10$&---&$9.6\times10^{-1}$&$6.2\times10^{-5}$&$       8.2$\\
\hline
\label{Tsummary}\end{tabular}}
Note: Dotted lines separate regime I from regime II runs.
Bracketed $k_\lambda$ values and hyphens mean that $k_\lambda$ exceeds $k_\mu$.
\end{table*}

In view of any type of driven or decaying MHD turbulence,
the dependence of $\EEGW^{\rm sat}$ on $\eta$ seems not
very intuitive as we find it to {\em increase} with increasing $\eta$
although one would have expected that {\em smaller} $\eta$ would cause
a more vigorous time dependence.
However, the increase of $\EEGW^{\rm sat}$ with $\eta$ is plausible due to the
fact that the maximum growth rate of $\BB$ is proportional to $\eta$
-- a specific of the CME.

In all of our simulations of series A--D, the parameter
$q$ is even lower than in the least efficient simulations of \cite{RPMBKK20}.
This is rather surprising and might indicate that the turbulence from
the CME has a much less vigorous time dependence than the cases
considered there.
For understanding the reason behind this, it is necessary to study
the present results in more detail by inspecting the magnetic and GW
energy spectra.
We begin by analyzing their mutual relation at late times
when the magnetic energy has already reached a maximum and
the GW energy has achieved a steady state.

\subsection{Late time GW spectra from the CME}

We consider the case $\eta=10^{-6}$, $\lambda=4\times10^8$, and
$\muz=10^4$, which corresponds to Run~B1.
This means that $\vmu=0.01$ and $\vlam=0.5$, so $\vlam/\vmu=50$,
and we are clearly in regime~I.

\begin{figure}\begin{center}
\includegraphics[width=\columnwidth]{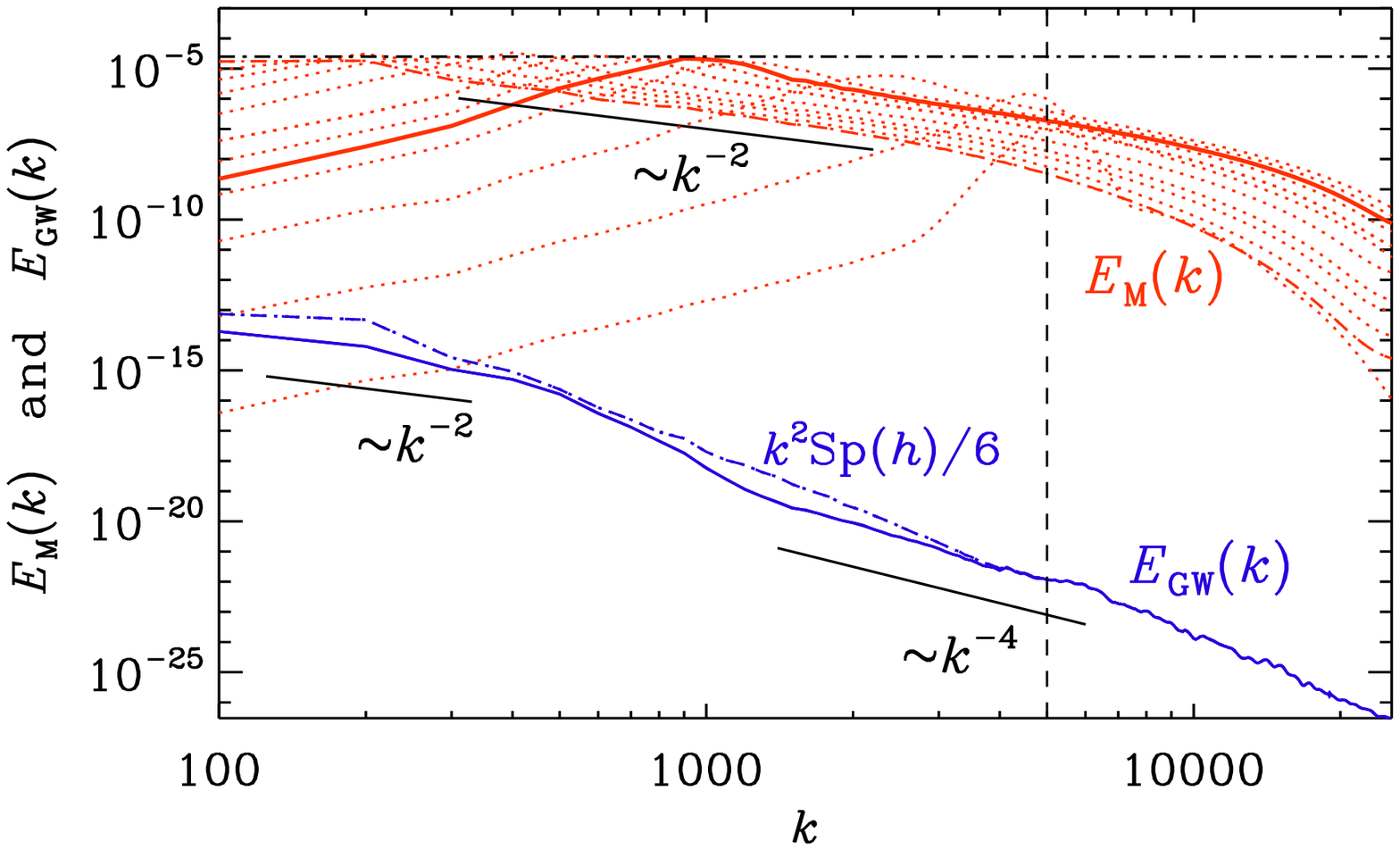}
\end{center}\caption{
Magnetic and GW energy spectra for Run~B1 with $\muz=10^4$,
$\lambda=4\times10^8$, and $\eta=10^{-6}$, which is in regime~I with
$(\vmu=0.01) < (\vlam=0.5)$.
$\EM(k)$ (red) is shown at the time of magnetic maximum (solid, $t=1.92$),
the time when the $k^{-2}$ spectrum is most clear ($t=3$, dashed),
and at selected other times (dotted, $t=1.71$, $1.77$, $1.83$, $1.89$, $1.94$, $2.00$, $2.15$,
$2.32$, $2.52$, $2.74$, and $3.00$, while $\EGW(k)$ (solid blue)
is from the simulation's end time ($t=14$), when it can be approximated
by $k^2\Sp(\hh)/6$ (dashed-dotted blue).
The black horizontal dashed-dotted line marks the saturation limit of \Eq{B2xiM},
$\muz/\lambda$, and the vertical dashed line marks the
position of $\kmu$.
}\label{pspec_sat_512_1e2_1e4_4e8_1em6aD}\end{figure}

The CME leads to exponential magnetic field generation, followed by
subsequent turbulent decay.
At the time of the magnetic maximum, an approximate $k^{-2}$ magnetic
energy spectrum with a short inertial range develops \citep{BSRKBFRK17}.
We then expect a $k^{-4}$ spectrum for the GW energy and a $k^{-6}$
spectrum for $\hh$; see \cite{RPMBKK20}.
There is a trend for this to happen also in the present case, although
$\EEGW(k)$ does not have clear power law subranges; see 
\Fig{pspec_sat_512_1e2_1e4_4e8_1em6aD}.
This is because the turbulence is not steady and both energy spectra look
very different even just shortly before the magnetic field saturates,
as will be shown below.

\begin{figure}\begin{center}
\includegraphics[width=\columnwidth]{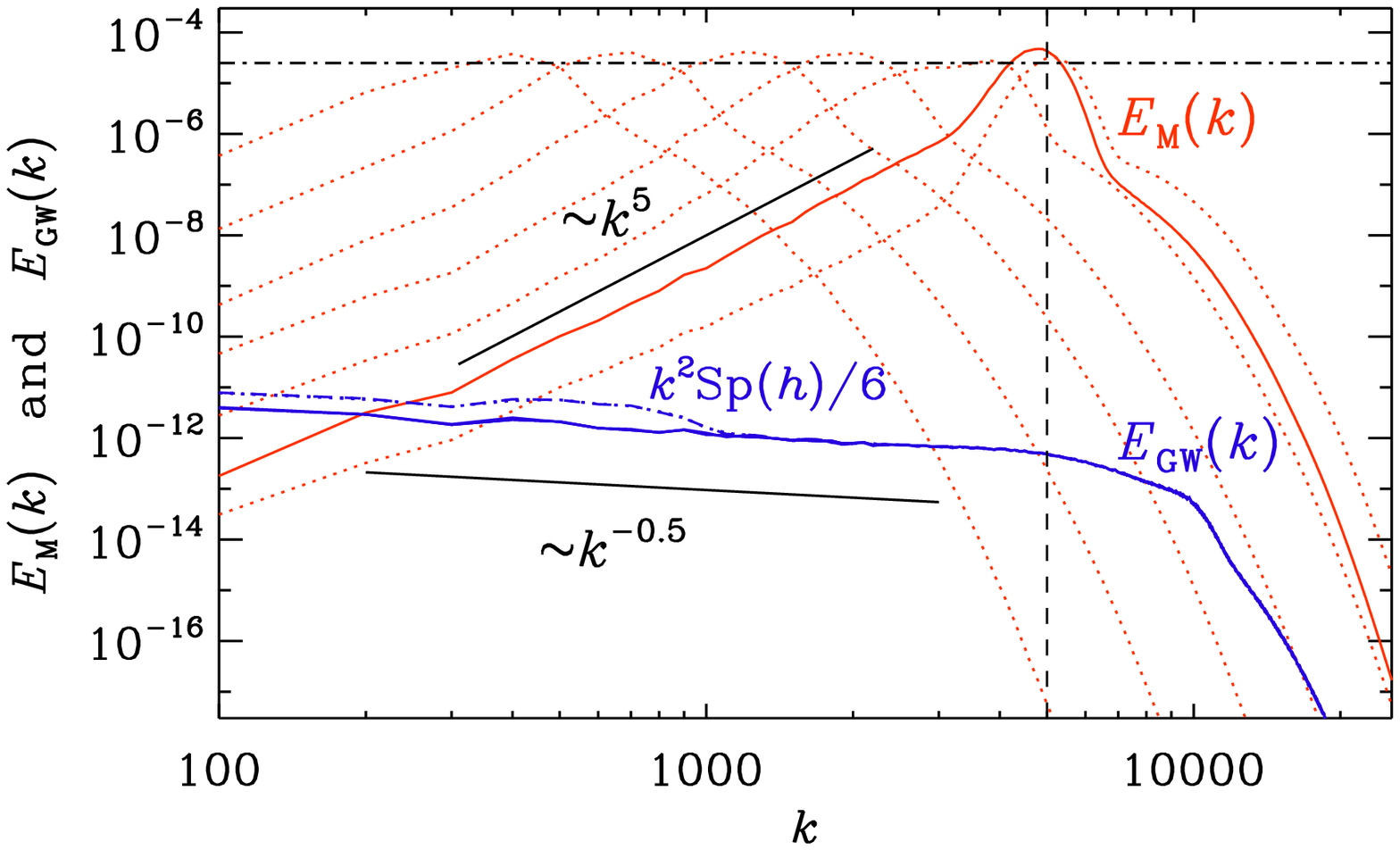}
\end{center}\caption{
Similar to \Fig{pspec_sat_512_1e2_1e4_4e8_1em6aD}, but for Run~B10 with
$\eta=10^{-3}$, which is in regime~II with $(\vmu=10) > (\vlam=0.5)$.
$E_{\rm M}(k)$ (solid red) is at the time when the magnetic energy has
attained its maximum ($t=1.001$), dotted red lines show $E_{\rm M}(k)$
at $t=1.0008$, $1.003$, $1.008$, $1.024$, and $1.075$,
while $\EGW(k)$ is from the simulation's end time
($t=1.075$), when $\EGW(k)\approx k^2\Sp(\hh)/6$.
}\label{pspec_sat_512_1e2_1e4_4e8_1em3a}\end{figure}

For runs in regime~II, however, we find an approximate $k^{-0.5}$ 
profile for $\EGW(k)$; see \Fig{pspec_sat_512_1e2_1e4_4e8_1em3a}.
This is closer to the case of stationary turbulence; see 
\Tab{Tregimes} for a comparison of some characteristic properties.
$\EM(k)$ shows an approximate $k^5$ subinertial range.
This is steeper than the $k^4$ spectrum expected based on causality
arguments \citep{DC03}.
However, as we will see later more clearly, at early times and
close to $k=\kmu$ the magnetic energy spectra show a dent,
explaining therefore the apparent steeper spectrum at early times;
a $k^4$ subinertial range can still be identified at other times.
In particular, for fully helical magnetic fields, a $k^4$ spectrum
spectrum always emerges, regardless of the initial slope; see Figure~3(a)
of \cite{BK17}.

%512_1e2_1e4_4e8_1em6a & 512_1e2_1e4_4e8_1em6aD (Run B1, regime I)
%512_1e2_1e4_4e8_1em3a (Run B10, regime II)
%M512sig1_k6_ramp1b (stationary turbulence)
\begin{table}[t!]\caption{
Spectral properties of GWs in regimes~I, II, and in stationary turbulence.
}\vspace{2pt}\begin{tabular}{lccccccc}
& Run~B1   & Run~B10   &  Stationary              \\
& Regime~I & Regime~II & $\;\;$turbulence$^*$\\
\hline
\vspace{-2mm}
\\
\vspace{+1mm}
$\Sp(\dot{\hh})/\Sp(\hh)$ & $0.89\,(2\gamma_0)^2$        & $0.96\,(2\gamma_0)^2$  & $0.30\,\kf^2$ \\
\vspace{+1mm}
$\Sp(\TT)/\Sp(\dot{\hh})$ & $1.1\,(2\gamma_0\muz)^2$ & $0.98\,(2\gamma_0)^2$  & $0.10\,\kf^2$ \\
\vspace{+1mm}
$\Sp(\hh)$, kinematic & $k^{-2}$ & $k^{2}$  & --- \\
\vspace{+1mm}
$\Sp(\hh)$, saturated & $k^{-2}$ & $k^{-0.5}$ & $k^0$ \\
\vspace{-6mm}\\
\label{Tregimes}\end{tabular}
\tablenotemark{
$^*$Run~K0 of \cite{Kahn21}, $\kf=600$; the ellipsis means no growth.}
\end{table}

\begin{figure}\begin{center}
\includegraphics[width=\columnwidth]{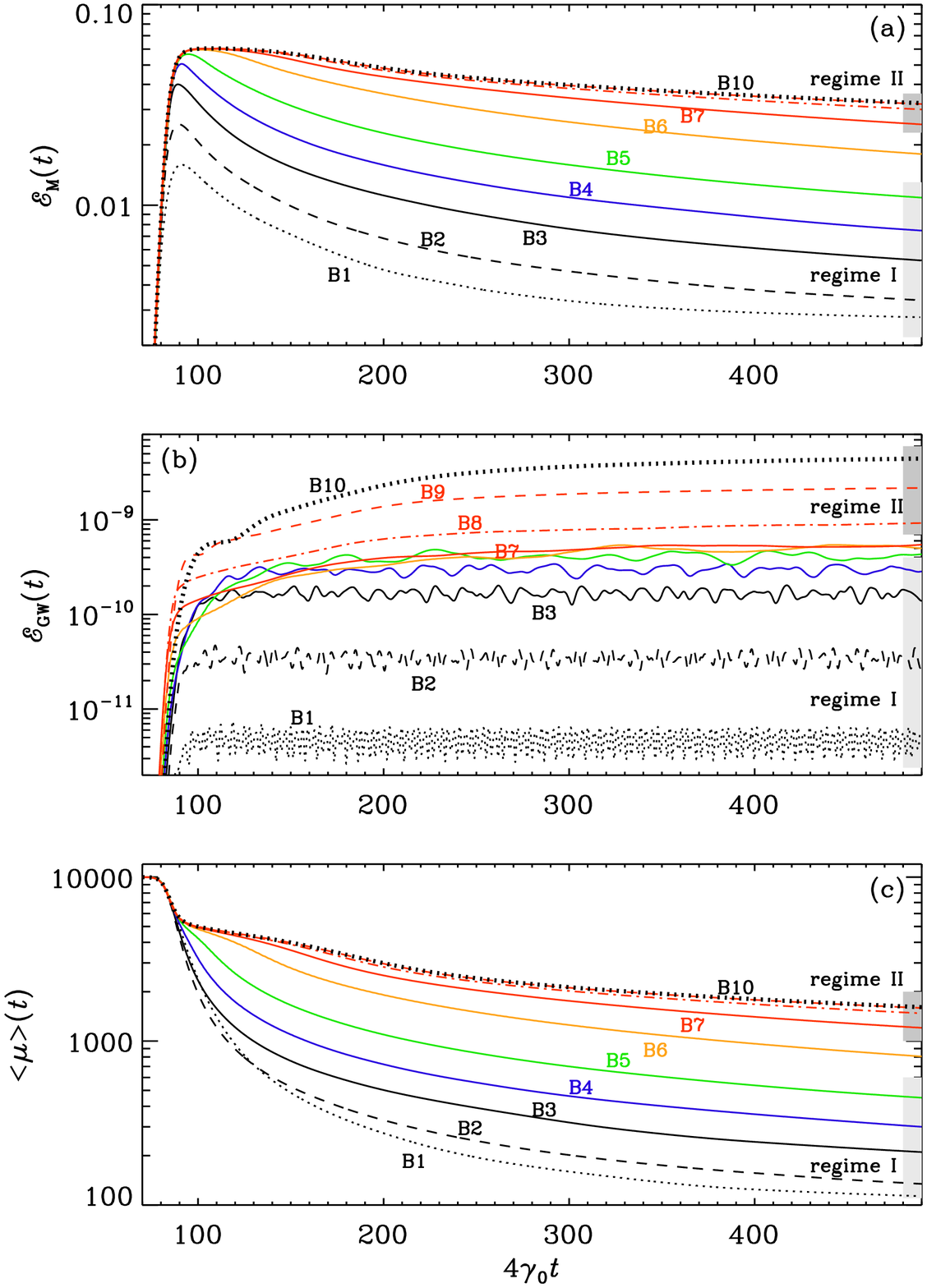}
\end{center}\caption{
Evolution of $\EEM$, $\EEGW$, and $\bra{\muf}$
for Runs~B1--B10.
The light and dark gray bars on the right of each panel indicate
regimes~I and II, respectively.
Note the occurrence of a slow final saturation phase of $\EEGW$
for all runs in regime~II (Runs~B7--B10).
Run~8 (red dashed-dotted line), Run~9 (red dashed line), and
Run~10 (upper black dotted line) overlap in $\EEM$ and
$\bra{\muf}$, but are well separated in $\EEGW$.
}\label{pcomp_seriesB}\end{figure}

\subsection{GW spectra during the early growth phase}
\label{EarlyPhase}

At early times, as discussed above, $\EEM(t)$ grows exponentially at a rate 
$2\gamma_0=\eta\muz^2/2$ and $\EEGW(t)$ grows at a rate $4\gamma_0$.
Across the different runs, this rate varies by three orders of magnitude.
To compare the evolution of GW and magnetic energies for the different
runs, it is thus convenient to plot both quantities versus $4\gamma_0 t$.
The result is shown in \Fig{pcomp_seriesB} for the runs of series~B.
One clearly sees a slow final saturation phase of $\EEGW(t)$
for all runs in regime~II (Runs~B7--B10), while $\EEM(t)$
and $\bra{\muf}(t)$ are almost unchanged across different runs.
During the exponential growth phase, $\muf$ is close to its initial
value, $\muz=10^4$.
It drops fastest in regime~I, where $\eta$ is small (Runs~B1--B5).
However, in contrast to \Fig{EEGW_vs_EEKM}, where we saw a marked
qualitative change as we move from regime~I to regime~II,
no such change is seen in \Fig{pcomp_seriesB} between regime~I
(Runs~B1--B5) and II (Runs~B7--B10).

In the case of stationary GW spectra \cite[see, e.g.,][]{Kahn21}, and also in
the previous section, we always have $\Sp(\dot{\hh})\approx k^2\Sp(\hh)\approx
k^{-2}\Sp(\TT)$, but this is not so in the early exponential growth phase.
Nevertheless, in both these regimes, we find $\Sp(\TT)\propto k^2$.
This is a consequence of the almost monochromatic magnetic field generation
in a narrow range around $k=\kmu$, which implies that the spectral
slope of $\EM(k)$ for $k<\kmu$ is always steeper than that of white noise
($\propto k^2$), so we call it ``blue noise''.
However, the square of a field with a blue noise spectrum always has a
white noise spectrum \citep{BB20}.
This explains why $\Sp(\TT)\propto k^2$.

\begin{figure}\begin{center}
\includegraphics[width=\columnwidth]{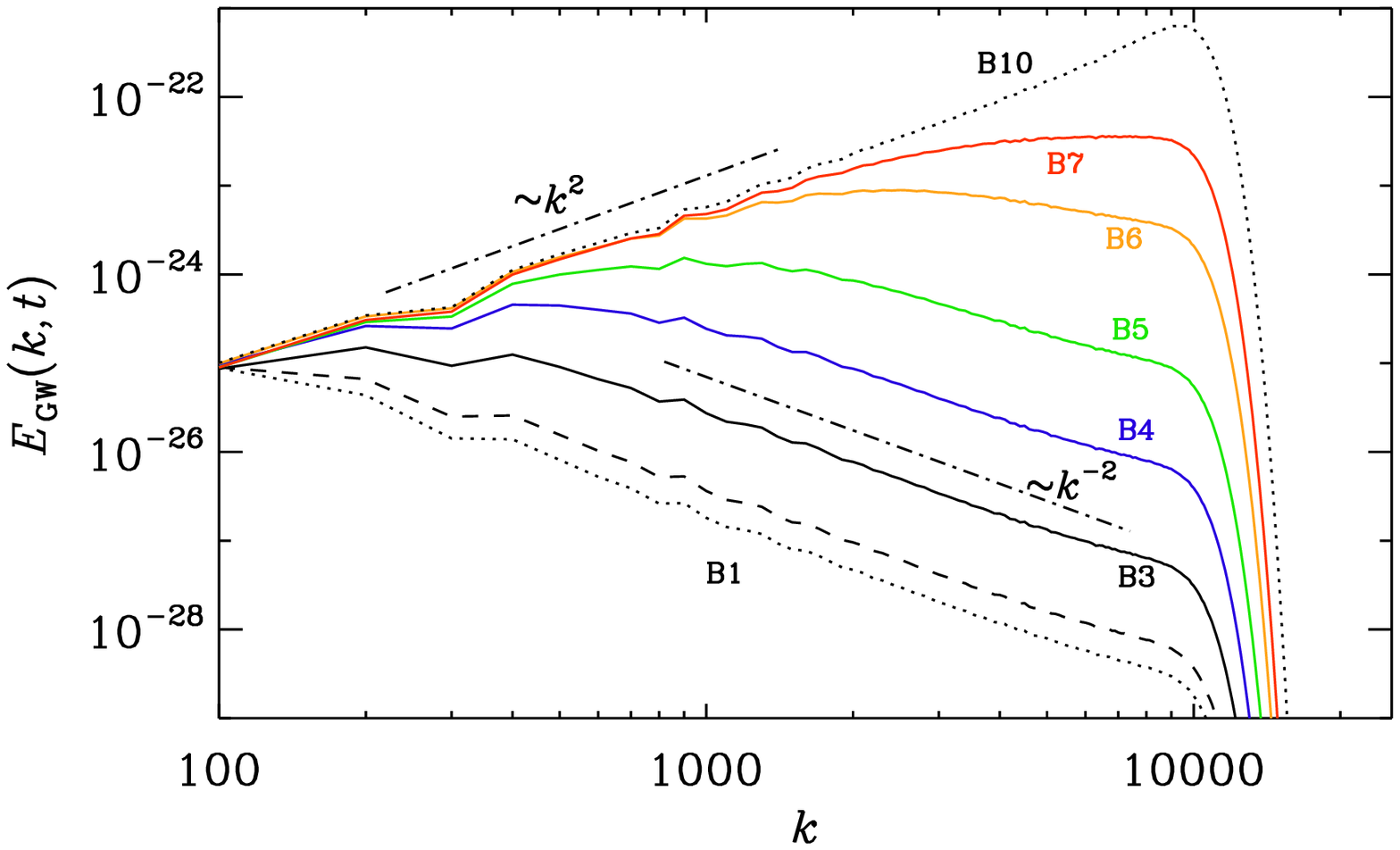}
\end{center}\caption{
Comparison of GW energy spectra during the kinematic growth stage
for runs in regime~I (B1--B5) and regime~II (B7--B10).
Note the change of slope at a certain wavenumber
that increases as we go from regime~I to regime~II.
}\label{ppspec_save}\end{figure}

\begin{figure*}\begin{center}
\includegraphics[width=\textwidth]{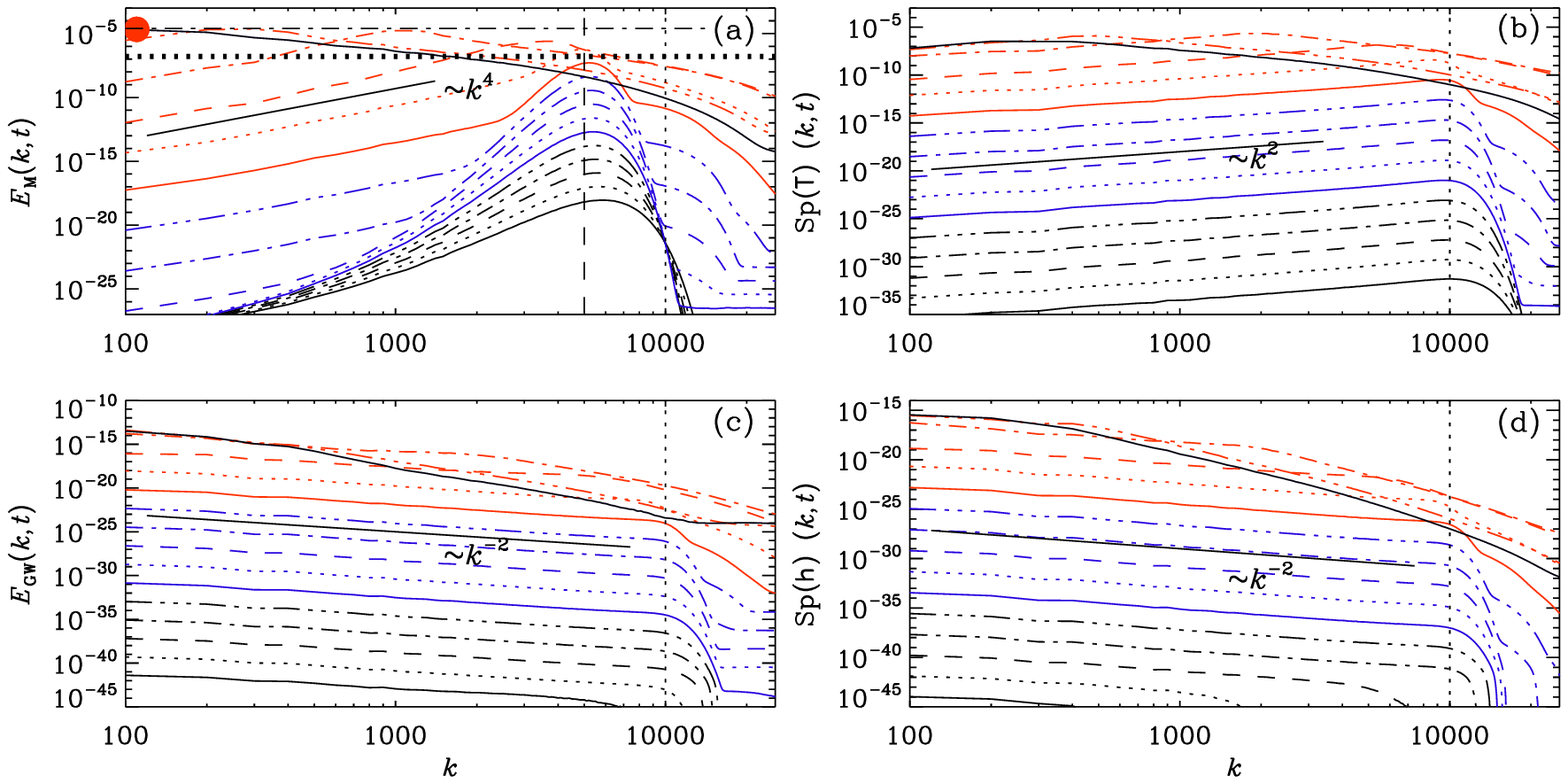}  
\end{center}\caption{
Time-evolving magnetic and GW energy spectra along with spectra
of stress $\TT$ and strain $\hh$ for Run~B1 (regime~I) at
$t-1=0.2$, $0.25$, $0.3$, $0.35$, $0.4$ in black,
$0.45$, $0.5$, $0.55$, $0.6$, $0.65$ in blue,
$0.7$, $0.75$, $0.8$, $0.9$, $1.4$ in red,
and the time of maximum $\EEM$ at $1.9$, again in black.
In panel (a), the dotted horizontal line marks the level of $C_5\muz\eta^2$,
and the horizontal dashed-dotted line the level of $C_\lambda\muz/\lambda$.
Vertical dotted and dashed lines mark the positions of $2\kmu=\muz$ 
and $\kmu$, respectively.
The red filled symbol denotes the peak of $\EM(k)$
at the time of the magnetic maximum.}
\label{pspec_select_512_1e2_1e4_4e8_1em6a}\end{figure*}

\begin{figure*}\begin{center}
\includegraphics[width=\textwidth]{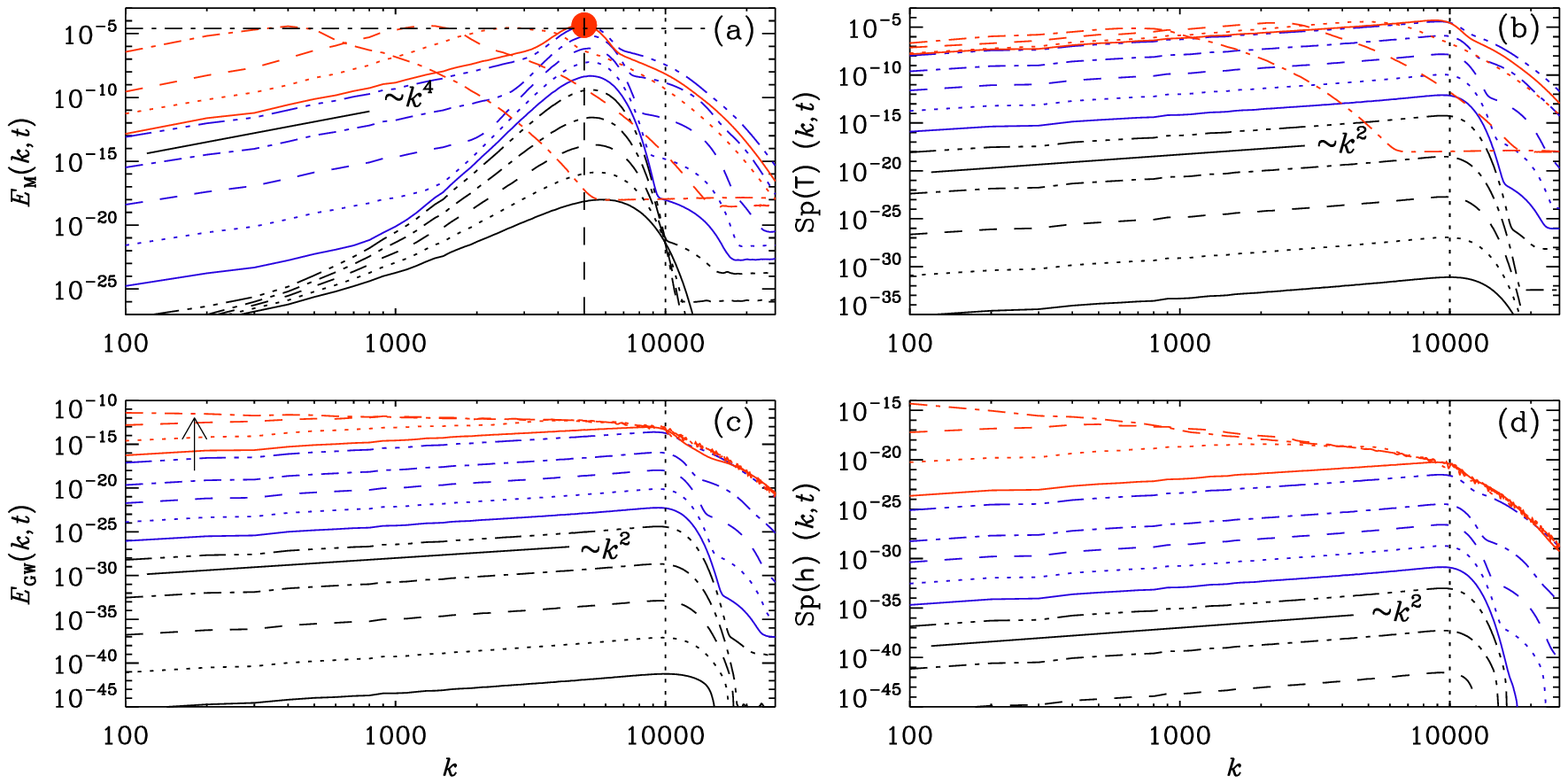}
\end{center}\caption{
Similar to \Fig{pspec_select_512_1e2_1e4_4e8_1em6a}, but
for Run~B10 with $\eta=10^{-3}$, which is in regime~II with
$(\vmu=10) > (\vlam=0.5)$, at
$t-1=2$, $3$, $4$, $5$, $6\times10^{-4}$ in black,
$6.5$, $7$, $7.5$, $8$, $9\times10^{-4}$ in blue, and
$0.001$ (maximum $\EEM$), $0.002$, $0.007$, and $0.0075$ in red.
$\EEM(t)$ reaches a maximum at $t-1=1.1\times10^{-3}$ and $\EEGW(t)$ at $t-1=0.02$.
The upward arrow in panel (c) emphasizes the change in slope.
}\label{pspec_select_512_1e2_1e4_4e8_1em3a}\end{figure*}

To see how the transition from a $k^{-2}$ profile for small $k$ toward
a $k^{2}$ profile for large $k$ occurs in $\EGW(k)$, we plot it in
\Fig{ppspec_save} during the kinematic growth phase.
The times have been arranged such that all spectra coincide at
$k=k_1\equiv100$.
We clearly see the emergence of a breakpoint from a $k^{2}$ spectrum
at low $k$ toward a $k^{-2}$ spectrum at larger $k$.  
The breakpoint shifts toward larger wavenumbers as we go from regime~I
to regime~II, although it can no longer be identified for Runs~B7--B10.

Furthermore, in both regimes~I and II, we find $\Sp(\dot{\hh})\propto\Sp(\hh)$
during the early growth phase, but their slopes are different in the two regimes.
In \Figs{pspec_select_512_1e2_1e4_4e8_1em6a}
{pspec_select_512_1e2_1e4_4e8_1em3a}, we compare the spectra for Runs~B1
(regime~I) and B10 (regime~II), including magnetic and GW energy spectra
along with the spectra of stress and strain.
We clearly see that at early times, $\Sp(\hh)$ and
$\Sp(\dot{\hh})=6\EEGW(k)$ all have the same slope proportional
to $k^{-2}$ and $k^2$ in regimes~I and II, respectively.
Specifically, at $k=\kmu$, we find for the ratio
$\Sp(\dot{\hh})/\Sp(\hh)\approx(2\gamma_0)^2$ in both regimes.
It is important to emphasize that, even though $\gamma(k)$ depends on $k$,
the stress spectrum grows at the {\em maximum} rate $\gamma_0$ at all $k$.
For $k\leq\kmu$, this can simply be understood as a consequence of
the result of \cite{BB20} that the square of a field with a blue noise
spectrum always has a white noise spectrum.

For $k>\muz$, the magnetic energy spectrum always drops rapidly.
Based again on the results of \cite{BB20}, since the spectrum is here
a red one, the magnetic stress spectrum also drops rapidly with the
same slope. 
Following \cite{BB20}, in the range $\kmu<k<\muz$ the spectrum is slightly
shallower than $k^2$ and it peaks approximately at $k=\muz$.

In \Tab{Tregimes}, we summarize the spectral properties during the early
kinematic growth phase and contrast it with the saturated phase.
In regime~I, we also find $\Sp(\TT)/\Sp(\dot{\hh})\approx(2\gamma_0)^2$, but
in regime~II, there is an extra $\muz^2$ factor (see \Tab{Tregimes}),
which is a consequence of the different slopes of both curves.
The reason for the change of slopes in regimes~I and II is explained
in the next section.

\subsection{Difference in the slopes between regimes~I and II}

To understand the change in the spectral slopes between regimes~I and
II during the kinematic growth stage it is convenient to restrict our
attention to the case of a purely monochromatic exponential growth of
$\BB$ at the wavenumber $\kmu$ with the rate $\gamma_0 = \eta\muz^2/4$.
As explained in \Sec{EarlyPhase}, the magnetic stress increases then at
all $k$ at the rate $2\gamma_0$; see also 
\Figs{pspec_select_512_1e2_1e4_4e8_1em6a}{pspec_select_512_1e2_1e4_4e8_1em3a}
for a direct confirmation of this property.

Let us now assume that $\tilde{T}(\kk,t)$, representing the
Fourier transform of one of the two polarization modes of the stress,
$T_+$ and $T_\times$, is given by
\EQ
\tilde{T}(\kk,t)=\theta(t-1)\,\tilde{T}_0(k)\,e^{2\gamma_0 (t-1)},
\EN
where $\theta(t)$ is the Heaviside step function, and $\tilde{T}_0(k)$
is assumed to depend just on $k=|\kk|$.

Using $\tilde{h}(k,1)=\dot{\tilde{h}}(k,1)=0$ as initial conditions,
we can solve \Eq{GW4} during the early growth phase in closed form as
\begin{align}
 \hspace*{-7mm}\tilde{h}(k,t) =
\frac{6\tilde{T}_0(k)}{4\gamma_0^2+k^2} 
\left[e^{2\gamma_0 \tau} -\cos k\tau
 -\frac{2\gamma_0}{k}\sin k\tau \right]_{\tau=t-1} \hspace{-8mm},  \hspace{-3mm}
 \label{specmod}
\end{align}
where $\tilde{h}$ stands for either $\tilde{h}_+$ or $\tilde{h}_\times$.
In practice, we are always interested in the case where the
exponential term dominates over the cosine and sine terms.
When $k\ll2\gamma_0$, $\Sp(h)$ and $\Sp(\dot{h})$ are proportional to $\Sp(T_0)$.
In particular, when $\tilde{T}_0(k)$ is a white noise spectrum, we have
$\Sp(h)\propto k^2|\tilde{T}_0(k)|^2\propto k^2$.
However, when $k\gg2\gamma_0$, we find
$\Sp(h)\propto\Sp(\dot{h})\propto k^2|\tilde{T}_0(k)/k^2|^2\propto k^{-2}$,
with the breakpoint being at $k_0=2\gamma_0$.

\begin{figure}\begin{center}
\includegraphics[width=.84\columnwidth]{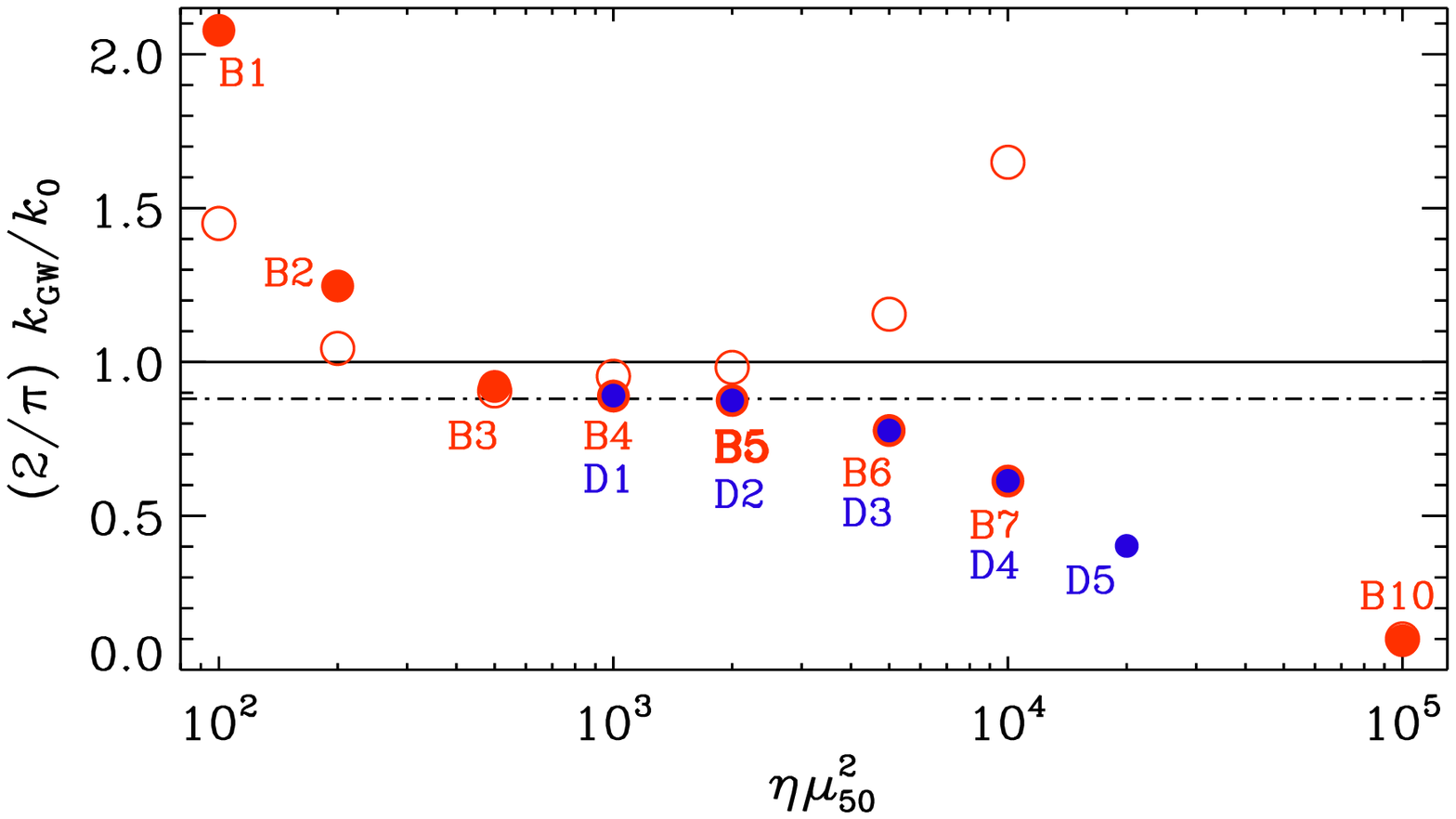}
\end{center}\caption{
Dependence of $k_{\rm GW}$ from \Eq{kGW}, normalized by $k_0\pi/2$,
on $\eta\muz^2$ for runs of series B (red filled symbols) and D (blue
filled symbols).
Run~B5 is highlighted in boldface (cf.\ \Fig{ppspec_save_B5}).
The dashed-dotted line gives an approximate fit through the data points
near their plateau, and the solid line goes through unity, the theoretically
expected value.
The red open symbols denote the values of $k_0$ obtained by fitting
the spectra of \Fig{ppspec_save} to the model spectrum of \Eq{EGWtheo},
similar to what is done in \Fig{ppspec_save_B5}.
}\label{pkGW_seriesB}\end{figure}

To compare with the results of our simulations, let us try to numerically
determine the breakpoint $k=k_{\rm GW}$ as       
\EQ
k_{\rm GW}^{-1}=\int k^{-1} E_{\rm GW}(k)\,\dd k\,
\left/ \int E_{\rm GW}(k)\,\dd k\right..
\label{kGW}
\EN
We have calculated it for the models of series~B and D and find that
our analytic prediction $k_0=2\gamma_0$ matches the numerical results
rather well; see \Fig{pkGW_seriesB}.
Representing $\Sp(\dot{h})$ according to \Eq{specmod} by
\EQ
E_{\rm GW}^{\rm model}\propto\left[\frac{k}{k_0^2+k^2}\,
e^{-(k/\muz)^4}\right]^2,
\label{EGWtheo}
\EN
where the exponential factor is intended to model the cutoff near
$k=\muz$, we find $k_{\rm GW}=(\pi/2)\,k_0$, which is why we have
compensated $k_{\rm GW}$ in \Fig{pkGW_seriesB} by this value.
The reason why there are departures for small and large values of
$\eta\muz^2$ is that the wavenumber range used for the integration
is limited. In addition to estimating $k_0$ as $k_{\rm GW}$ from \Eq{kGW},
we compute a fit to the model spectrum of \Eq{EGWtheo}.
We do this by minimizing the mean squared difference between the
actual spectrum and the model spectrum.
Those results are also shown in \Fig{pkGW_seriesB} (open symbols).

\begin{figure}\begin{center}
\includegraphics[width=.88\columnwidth]{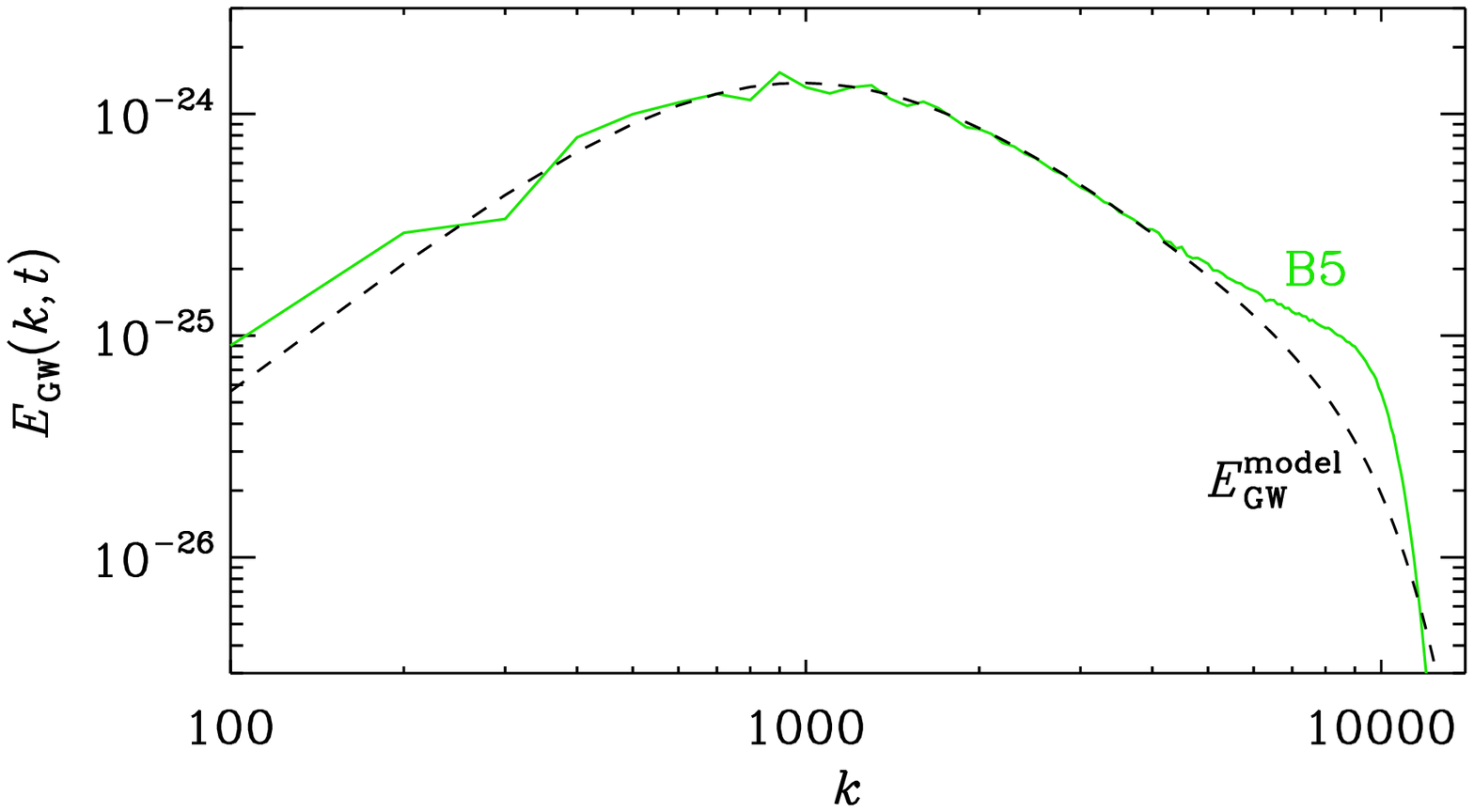}
\end{center}\caption{
Comparison of the GW energy spectrum for Run~B5 and the model spectrum
of \Eq{EGWtheo}.
}\label{ppspec_save_B5}\end{figure}

\begin{figure*}\begin{center}
\includegraphics[width=\textwidth]{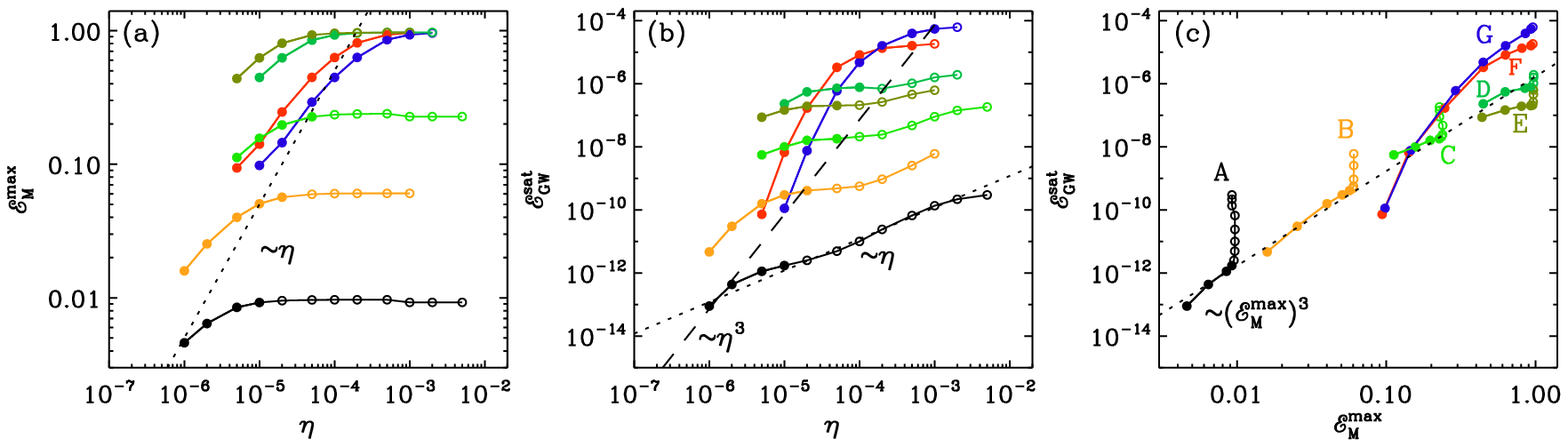}
\end{center}\caption{
Dependence of $\EEM^{\rm max}$ and $\EEGW^{\rm sat}$ on $\eta$, and their
mutual parametric dependence for runs of series~A--D with $\muz=10^4$
and $\lambda^{1/2}=5\times10^4$, $2\times10^4$, $10^4$, $5\times10^3$, respectively,
series~E with $\muz=2\times10^4$, $\lambda^{1/2}=10^4$, 
series~F with $\muz=2000$, $\lambda^{1/2}=1000$, and
series~G with $\muz=1000$, $\lambda^{1/2}=100$.
Filled (open) symbols denote runs in regime I (II).
The dotted line in panel (c) is for $q=13\,(\EEM^{\rm max})^{1/2}$.
}\label{peta}\end{figure*}

In \Fig{ppspec_save_B5}, we show a comparison 
of one of the GW energy spectra of \Fig{ppspec_save} with \Eq{EGWtheo}.
While it provides an excellent description of $\EGW(k)$ in the bulk of
the $k$ range, the exponential factor is not sharp enough to model the
simulation data near the cutoff.

\subsection{Change of slope toward late times}
\label{LateTimes}

We see in \Fig{pcomp_seriesB}(b) that for all runs in regime~I, $\EEGW$ 
saturates quickly after
$\EEM$ reaches its maximum, while for runs in regime~II, $\EEGW$ continues
to display a slow saturation behavior.
To understand this unusual behavior, we must look again at
\Fig{pspec_select_512_1e2_1e4_4e8_1em3a}, showing the evolution 
of the spectra in Run~B10, which is in regime~II.
We see that, at the time when $\EEM$ reaches its maximum, the peak of
$\EM(k)$ is still at $k\approx\muf$.
After that, $\EEM$ decays such that $\EEM/\kM=\const$, so
based on the earlier results of \cite{RPMBKK20},
we would expect $\EEGW$ to stay constant.
Looking at the evolution of $\EGW(k)$ for Run~B10 near equilibration in
\Fig{pspec_select_512_1e2_1e4_4e8_1em3a}(c), we observe a change in slope.
This could be responsible for the occurrence of a slow final saturation
phase of $\EEGW$ for the runs in regime~II, and especially
for Run~B10, seen in \Fig{pcomp_seriesB}.

To discuss this possibility quantitatively, let us assume a
simplified spectrum of the form
\EQ
E_{\rm GW}(k,t_{\rm bef})=3{\cal E}_0 \,k^2/\muz^3
\quad\mbox{if $k<2\kmu=\muz$}  
\label{EGW_tbef}
\EN
for the time $t_{\rm bef}$ before the slope changes.
For $k>2\kmu$ we assume a sharp fall-off and 
therefore ignore that contribution.
This $k^2$ spectrum is normalized such that
$\int E_{\rm GW}(k,t_{\rm bef})\,\dd k={\cal E}_0$.
The spectrum is then assumed to change to a new power
law $\propto k^s$, with an exponent $s$, of the form
\EQ
E_{\rm GW}(k,t_{\rm aft})=3{\cal E}_0\,k^s/\muz^{s+1}
\label{EGW_taft}
\EN
for the time $t_{\rm aft}$ after the slope has changed.
Employing the same ${\cal E}_0$ in \Eqs{EGW_tbef}{EGW_taft}, accounts
for the fact that $E_{\rm GW}(\muz) = 3 {\cal E}_0/\muz $ is no longer
changing in time; see \Fig{pspec_select_512_1e2_1e4_4e8_1em3a}(c).
For $s>-1$, the resulting GW energy is $3{\cal E}_0/(s+1)$.
In \Fig{pspec_sat_512_1e2_1e4_4e8_1em3a}, we found $s=-0.5$, so the
resulting GW energy is then $\approx6{\cal E}_0$, which is compatible
with the late-time excess of $\EEGW$ in Run~B10 relative to Run~B6.
It should be noted, however, that the change of slope occurs at a time
when the magnetic field is about to reach the scale of the domain.
It is therefore conceivable that $s=-0.5$
could be an artifact of the finite domain size.
In particular, $s=0$ is what has previously been found based on
numerical simulations \citep{RPMBKK20} including larger domains.

\begin{figure*}\begin{center}
\includegraphics[width=\textwidth]{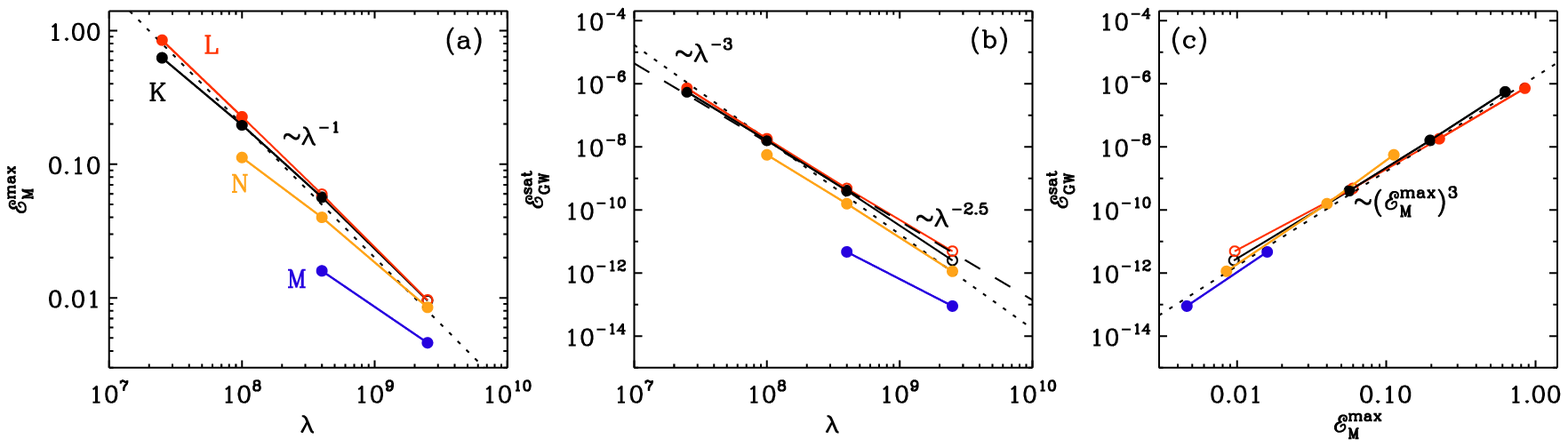}
\end{center}\caption{
Dependence of $\EEM^{\rm max}$ and $\EEGW^{\rm sat}$ on $\lambda$, and their
mutual parametric dependence for runs of series~K--N.
Filled (open) symbols denote runs in regime I (II).
The dotted line in panel (c) is for $q=7\,\EEM^{\max}$.
}\label{plam}\end{figure*}

\begin{figure*}\begin{center}
\includegraphics[width=\textwidth]{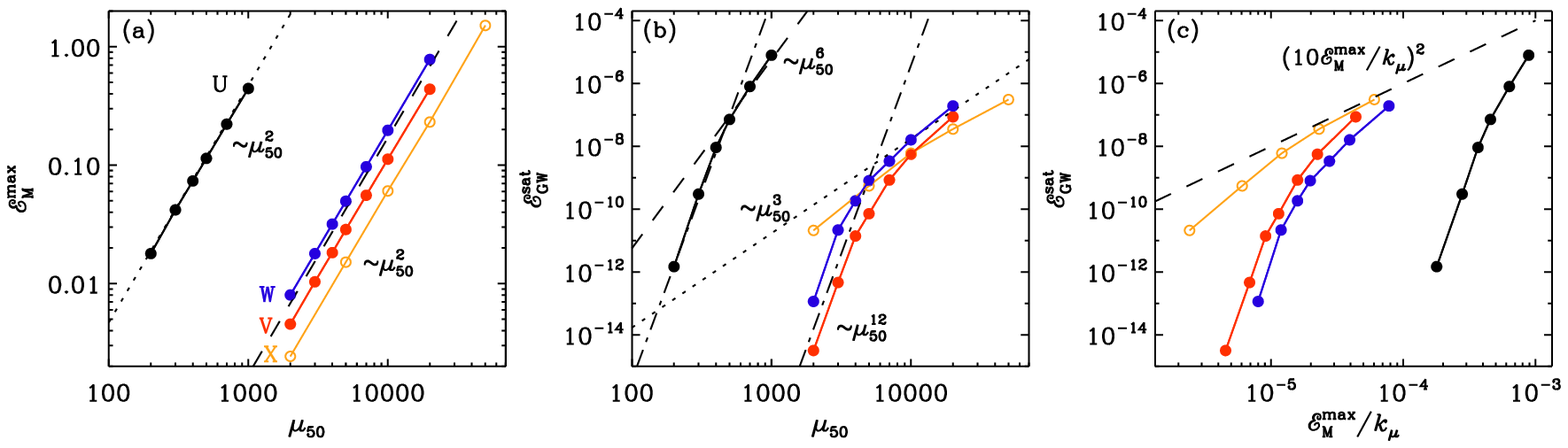}
\end{center}\caption{
Dependence of $\EEM^{\rm max}$ and $\EEGW^{\rm sat}$ on $\muz$, and their
mutual parametric dependence for runs of series~U--X.
Filled (open) symbols denote runs in regime I (II).
The dashed line in panel (c) is for $q=10$.
}\label{pmu}\end{figure*}

\begin{figure}\begin{center}
\includegraphics[width=.90\columnwidth]{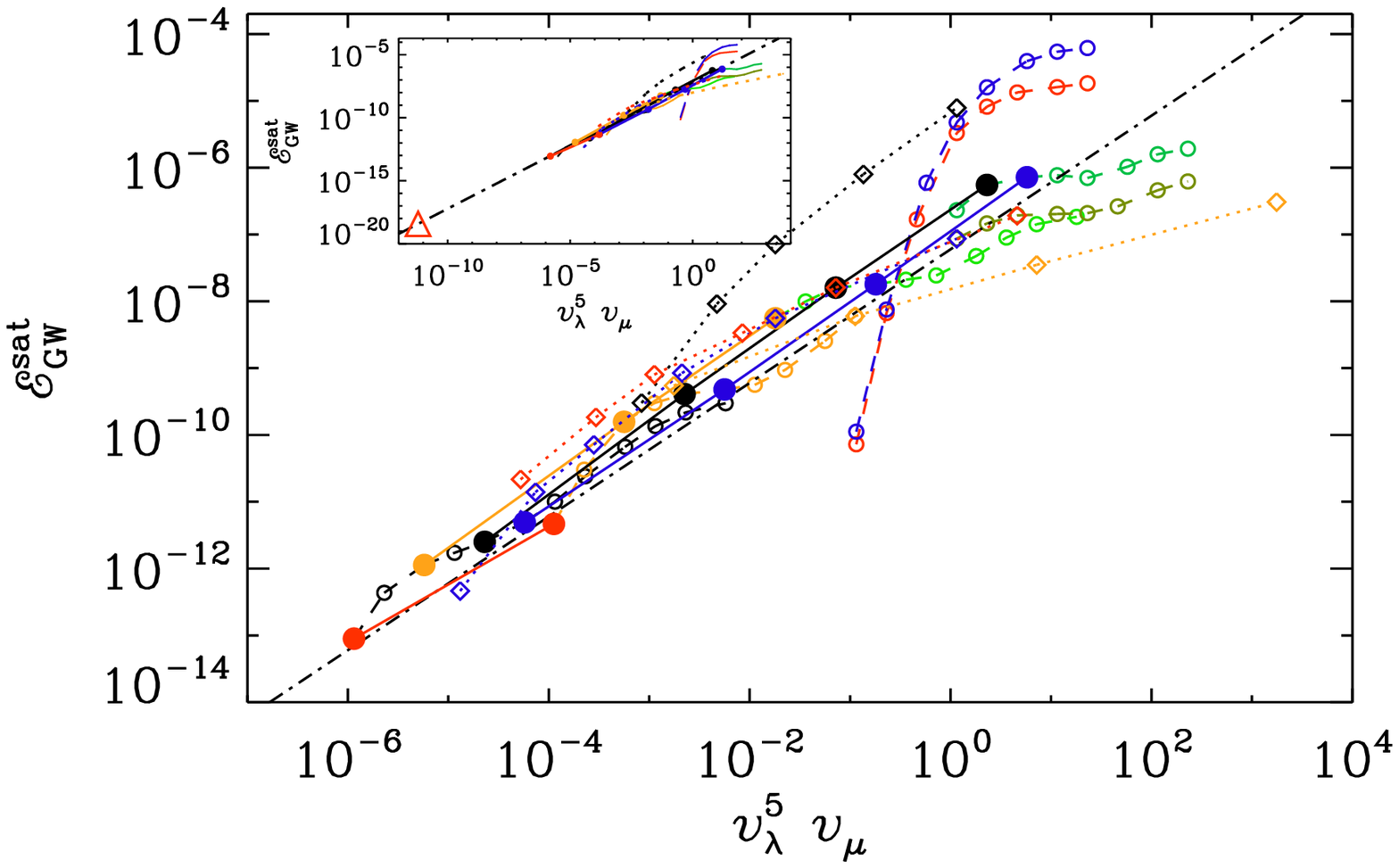}
\end{center}\caption{
Dependence of $\EEGW^{\rm sat}$ on $\vlam^5\vmu$ for all runs of
series~A--G with the same colors as in \Fig{peta} (dashed lines with open circles),
series~K--N with the same colors as in \Fig{plam} (solid lines with filled symbols), and
series~U--X with the same colors as in \Fig{pmu} (dotted lines with diamonds).
The dashed-dotted line has slope unity.
The inset shows the same plot, extended
down to $\vlam^5\vmu=6\times10^{-12}$, corresponding to the CME estimate
for the early universe, and denoted by a big red triangle in the lower left.
}\label{pall}\end{figure}

\begin{table*}\caption{
Summary of Runs from series K--N.
}\vspace{2pt}\centerline{\begin{tabular}{l@{\hspace{-0.5mm}}lccccccccccccc}
Run & &$\eta$ & $\lambda^{1/2}$ &$\muz$ &$\eta k_1$& $\vmu$ & $\vlam$ & $\eta\muz^2$ & $k_1$ & $k_\lambda$ & $\EEM^{\max}$ & $\EEGW^{\rm sat}$ & $q$ \\[.7mm]
%Run   eta      lam1/2     mu       etak1      vmu       vlam       k1       klam    EEM       EEGW   q
\hline\\[-3mm] 
K1&= A5&$2\times10^{-5}$&$5\times10^{4}$&$10^{4}$&$2\times10^{-3}$&$       0.2$&$       0.2$&$      2000$&$       100$&---&$9.5\times10^{-3}$&$2.5\times10^{-12}$&$       1.7$\\
K2&= B5&$2\times10^{-5}$&$2\times10^{4}$&$10^{4}$&$2\times10^{-3}$&$       0.2$&$       0.5$&$      2000$&$       100$&---&$5.7\times10^{-2}$&$4.1\times10^{-10}$&$       1.4$\\
K3&= C3&$2\times10^{-5}$&$1\times10^{4}$&$10^{4}$&$2\times10^{-3}$&$       0.2$&$         1$&$      2000$&$       100$&$      8000$&$2.0\times10^{-1}$&$1.6\times10^{-8}$&$       1.3$\\
K4&= D2&$2\times10^{-5}$&$5\times10^{3}$&$10^{4}$&$2\times10^{-3}$&$       0.2$&$         2$&$      2000$&$       100$&$      4000$&$6.3\times10^{-1}$&$5.5\times10^{-7}$&$       1.2$\\
           \hline 
L1&= A6&$5\times10^{-5}$&$5\times10^{4}$&$10^{4}$&$5\times10^{-3}$&$       0.5$&$       0.2$&$      5000$&$       100$&---&$9.6\times10^{-3}$&$4.9\times10^{-12}$&$       2.3$\\
L2&= B6&$5\times10^{-5}$&$2\times10^{4}$&$10^{4}$&$5\times10^{-3}$&$       0.5$&$       0.5$&$      5000$&$       100$&---&$6.0\times10^{-2}$&$4.8\times10^{-10}$&$       3.7$\\
L3&= C4&$5\times10^{-5}$&$1\times10^{4}$&$10^{4}$&$5\times10^{-3}$&$       0.5$&$         1$&$      5000$&$       100$&---&$2.3\times10^{-1}$&$1.8\times10^{-8}$&$       3.0$\\
L4&= D3&$5\times10^{-5}$&$5\times10^{3}$&$10^{4}$&$5\times10^{-3}$&$       0.5$&$         2$&$      5000$&$       100$&$     10000$&$8.5\times10^{-1}$&$7.2\times10^{-7}$&$       2.5$\\
           \hline
M1&= A1&$1\times10^{-6}$&$5\times10^{4}$&$10^{4}$&$1\times10^{-4}$&$      0.01$&$       0.2$&$       100$&$       100$&$      2000$&$4.6\times10^{-3}$&$8.9\times10^{-14}$&$     0.032$\\
M2&= B1&$1\times10^{-6}$&$2\times10^{4}$&$10^{4}$&$1\times10^{-4}$&$      0.01$&$       0.5$&$       100$&$       100$&$       800$&$1.6\times10^{-2}$&$4.7\times10^{-12}$&$     0.027$\\
           \hline
N1&= A3&$5\times10^{-6}$&$5\times10^{4}$&$10^{4}$&$5\times10^{-4}$&$      0.05$&$       0.2$&$       500$&$       100$&$     10000$&$8.5\times10^{-3}$&$1.1\times10^{-12}$&$      0.31$\\
N2&= B3&$5\times10^{-6}$&$2\times10^{4}$&$10^{4}$&$5\times10^{-4}$&$      0.05$&$       0.5$&$       500$&$       100$&$      4000$&$4.0\times10^{-2}$&$1.6\times10^{-10}$&$      0.31$\\
N3&= C1&$5\times10^{-6}$&$1\times10^{4}$&$10^{4}$&$5\times10^{-4}$&$      0.05$&$         1$&$       500$&$       100$&$      2000$&$1.1\times10^{-1}$&$5.6\times10^{-9}$&$      0.33$\\
\hline
\label{Tsummary_lam}\end{tabular}}\end{table*}

\begin{table*}\caption{
Summary of Runs from series U--X.
}\vspace{2pt}\centerline{\begin{tabular}{lccccccccccccc}
Run &$\eta$ & $\lambda^{1/2}$ &$\muz$ &$\eta k_1$& $\vmu$ & $\vlam$ & $\eta\muz^2$ & $k_1$ & $k_\lambda$ & $\EEM^{\max}$ & $\EEGW^{\rm sat}$ & $q$ \\[.5mm]
%Run   eta      lam1/2     mu       etak1      vmu       vlam       k1       klam    EEM       EEGW   q
\hline
U1&$1\times10^{-4}$&$5\times10^{2}$&$2\times10^{2}$&$1\times10^{-3}$&$     0.020$&$       0.4$&$         4$&$        10$&$        40$&$1.8\times10^{-2}$&$1.5\times10^{-12}$&$   0.00068$\\
U2&$1\times10^{-4}$&$5\times10^{2}$&$3\times10^{2}$&$1\times10^{-3}$&$     0.030$&$       0.6$&$         9$&$        10$&$        60$&$4.2\times10^{-2}$&$3.0\times10^{-10}$&$    0.0062$\\
U3&$1\times10^{-4}$&$5\times10^{2}$&$4\times10^{2}$&$1\times10^{-3}$&$     0.040$&$       0.8$&$        16$&$        10$&$        80$&$7.4\times10^{-2}$&$9.2\times10^{-9}$&$     0.026$\\
U4&$1\times10^{-4}$&$5\times10^{2}$&$5\times10^{2}$&$1\times10^{-3}$&$     0.050$&$         1$&$        25$&$        10$&$       100$&$1.1\times10^{-1}$&$7.1\times10^{-8}$&$     0.058$\\
U5&$1\times10^{-4}$&$5\times10^{2}$&$7\times10^{2}$&$1\times10^{-3}$&$     0.070$&$         1$&$        49$&$        10$&$       140$&$2.2\times10^{-1}$&$8.0\times10^{-7}$&$      0.14$\\
U6$\,=$G4&$1\times10^{-4}$&$5\times10^{2}$&$1\times10^{3}$&$1\times10^{-3}$&$      0.10$&$         2$&$       100$&$        10$&$       200$&$4.4\times10^{-1}$&$7.9\times10^{-6}$&$      0.32$\\
\hline
V1&$5\times10^{-6}$&$10^{4}$&$2\times10^{3}$&$5\times10^{-4}$&$     0.010$&$       0.2$&$        20$&$       100$&$       400$&$4.5\times10^{-3}$&$3.2\times10^{-15}$&$    0.0012$\\
V2&$5\times10^{-6}$&$10^{4}$&$3\times10^{3}$&$5\times10^{-4}$&$     0.015$&$       0.3$&$        45$&$       100$&$       600$&$1.0\times10^{-2}$&$4.6\times10^{-13}$&$     0.010$\\
V3&$5\times10^{-6}$&$10^{4}$&$4\times10^{3}$&$5\times10^{-4}$&$     0.020$&$       0.4$&$        80$&$       100$&$       800$&$1.8\times10^{-2}$&$1.4\times10^{-11}$&$     0.041$\\
V4&$5\times10^{-6}$&$10^{4}$&$5\times10^{3}$&$5\times10^{-4}$&$     0.025$&$       0.5$&$       125$&$       100$&$      1000$&$2.9\times10^{-2}$&$7.1\times10^{-11}$&$     0.074$\\
V5&$5\times10^{-6}$&$10^{4}$&$7\times10^{3}$&$5\times10^{-4}$&$     0.035$&$       0.7$&$       245$&$       100$&$      1400$&$5.6\times10^{-2}$&$8.4\times10^{-10}$&$      0.18$\\
V6$\,=$C1&$5\times10^{-6}$&$10^{4}$&$1\times10^{4}$&$5\times10^{-4}$&$     0.050$&$         1$&$       500$&$       100$&$      2000$&$1.1\times10^{-1}$&$5.6\times10^{-9}$&$      0.33$\\
V7&$5\times10^{-6}$&$10^{4}$&$2\times10^{4}$&$1\times10^{-3}$&$      0.10$&$         2$&$      2000$&$       200$&$      4000$&$4.4\times10^{-1}$&$8.6\times10^{-8}$&$      0.67$\\
\hline
W1&$2\times10^{-5}$&$10^{4}$&$2\times10^{3}$&$2\times10^{-3}$&$     0.040$&$       0.2$&$        80$&$       100$&$      1600$&$8.0\times10^{-3}$&$1.2\times10^{-13}$&$     0.017$\\
W2&$2\times10^{-5}$&$10^{4}$&$3\times10^{3}$&$2\times10^{-3}$&$     0.060$&$       0.3$&$       180$&$       100$&$      2400$&$1.8\times10^{-2}$&$2.2\times10^{-11}$&$      0.16$\\
W3&$2\times10^{-5}$&$10^{4}$&$4\times10^{3}$&$2\times10^{-3}$&$     0.080$&$       0.4$&$       320$&$       100$&$      3200$&$3.2\times10^{-2}$&$1.8\times10^{-10}$&$      0.34$\\
W4&$2\times10^{-5}$&$10^{4}$&$5\times10^{3}$&$2\times10^{-3}$&$      0.10$&$       0.5$&$       500$&$       100$&$      4000$&$5.0\times10^{-2}$&$8.0\times10^{-10}$&$      0.57$\\
W5&$2\times10^{-5}$&$10^{4}$&$7\times10^{3}$&$2\times10^{-3}$&$      0.14$&$       0.7$&$       980$&$       100$&$      5600$&$9.7\times10^{-2}$&$3.4\times10^{-9}$&$      0.84$\\
W6$\,=$C3&$2\times10^{-5}$&$10^{4}$&$1\times10^{4}$&$2\times10^{-3}$&$      0.20$&$         1$&$      2000$&$       100$&$      8000$&$2.0\times10^{-1}$&$1.6\times10^{-8}$&$       1.3$\\
W7&$2\times10^{-5}$&$10^{4}$&$2\times10^{4}$&$2\times10^{-3}$&$      0.40$&$         2$&$      8000$&$       100$&$     16000$&$7.8\times10^{-1}$&$1.9\times10^{-7}$&$       2.2$\\
\hline
X1&$1\times10^{-3}$&$2\times10^{4}$&$2\times10^{3}$&$1\times10^{-1}$&$       2.0$&$       0.1$&$      4000$&$       100$&---&$2.4\times10^{-3}$&$2.1\times10^{-11}$&$       3.8$\\
X2&$1\times10^{-3}$&$2\times10^{4}$&$5\times10^{3}$&$1\times10^{-1}$&$       5.0$&$       0.2$&$     25000$&$       100$&---&$1.5\times10^{-2}$&$5.5\times10^{-10}$&$       7.7$\\
X3$\,=$B10&$1\times10^{-3}$&$2\times10^{4}$&$1\times10^{4}$&$1\times10^{-1}$&$        10$&$       0.5$&$1\times10^{5}$&$       100$&---&$6.0\times10^{-2}$&$6.0\times10^{-9}$&$        12$\\
X4&$1\times10^{-3}$&$2\times10^{4}$&$2\times10^{4}$&$1\times10^{-1}$&$        20$&$         1$&$4\times10^{5}$&$       100$&---&$2.3\times10^{-1}$&$3.5\times10^{-8}$&$        16$\\
X5&$1\times10^{-3}$&$2\times10^{4}$&$5\times10^{4}$&$5\times10^{-1}$&$        50$&$         2$&$2\times10^{6}$&$       500$&---&$1.5\times10^{0}$&$3.1\times10^{-7}$&$        18$\\
\hline
\label{Tsummary_mu}\end{tabular}}\end{table*}

\subsection{Dependence on $\eta$ for given $\lambda$ and $\muz$}

We have already seen that, as we increase $\eta$, we gradually move from
regime~I to regime~II.
Let us now also determine the functional dependence of both
$\EEM^{\max}$ and $\EEGW^{\rm sat}$ on $\eta$.
This is shown in \Fig{peta} for the runs of series~A--G.

In \Fig{peta}(c), we see the dotted line describing a cubic dependence,
$\EEGW^{\rm sat}\approx1.7\times10^{-6}\,(\EEM^{\max})^3$.
A similar scaling has also been suggested by \cite{Neronov20} based on
the consideration of characteristic time and length scales.
Using \Eq{qdef2}, this implies a square root dependence of 
the efficiency parameter $q$, $q\approx13\,(\EEM^{\max})^{1/2}$.

As expected from \Eq{Bsat}, smaller values of $\lambda$ lead to an
increase of $\EEM^{\max}$.
Values close to unity become not only more unrealistic because of Big
Bang nucleosynthesis constraints \citep{GR01}, but they also 
can more easily lead to numerical problems.

In all cases, we see that there is a change in slope and that
$\EEM^{\max}$ reaches a plateau when
$\vmu/\vlam=\eta\lambda^{1/2}$ approaches a critical value of around one half.
Interestingly, $\EEGW^{\rm sat}$ still continues to increase
approximately linearly with $\eta$, so this cannot be explained by an
increase of $\EEM^{\max}$.
However, we have seen in \Sec{LateTimes} that there is
a change in the slope
of $E_{\rm GW}(k)$, which results in larger GW energy when the slope
changes from $k^2$ to $k^0$ or even $k^{-0.5}$; see \Eq{EGW_taft}.

\subsection{Dependence on $\lambda$ for given $\eta$ and $\muz$}
\label{LamDep}

As expected, $\EEM^{\max}$ scales inversely proportional
to $\lambda$.
This can be seen in the first panel of \Fig{plam}, where we plot the runs
of series~K--N; see also \Tab{Tsummary_lam} for a summary.
In the other panels, we also show the dependence of $\EEGW^{\rm sat}$
on $\lambda$ and the mutual parametric dependence of
$\EEGW^{\rm sat}$ on $\EEM^{\max}$.
We see that the dependence of $\EEGW^{\rm sat}$ on $\lambda$
is steeper than $\lambda^{-2}$
-- approximately like $\propto\lambda^{-5/2}$, according to \FFig{plam}(b).
The dependence of $\EEGW^{\rm sat}$ on $\EEM^{\max}$
is therefore also steeper than quadratic, namely approximately cubic;
see \FFig{plam}(c).

It is instructive to see how well the dependence of
$\EEGW^{\rm sat}$ on $\eta$, $\lambda$, and $\muz$
can be expressed just in terms of $\vlam$ and $\vmu$.
The approximately linear dependence of $\EEGW^{\rm sat}$ on
$\eta$ seen in \Fig{peta}(b) for regime~II would then also suggest 
its linear dependence on $\vmu$.
Furthermore, the approximate scaling
$\EEGW^{\rm sat}\propto\lambda^{-5/2}$ seen in \Fig{plam}(b)
would suggest $\EEGW^{\rm sat}\propto\vlam^5$.
The combined dependence would then be
\EQ
\EEGW^{\rm sat}\propto\vlam^5\vmu,
\label{vlam5vmu}
\EN
implying $\EEGW^{\rm sat}\propto\muz^6$.
In the next section we see that this suggestion agrees
reasonably well with our data.

\subsection{Dependence on $\muz$ for given $\eta$ and $\lambda$}

Let us finally determine the dependence of $\EEM^{\max}$
and $\EEGW^{\rm sat}$ on $\muz$, keeping $\eta$ and $\lambda$
unchanged.
The results are shown in \Fig{pmu}.
We clearly see the expected quadratic dependence of
$\EEM^{\max}$ on $\muz$.
The dependence of $\EEGW^{\rm sat}$ on $\muz$ is much steeper
and shows a break at $\muz\approx500$ for runs of series~U and
$5000$ for runs of series V and W.
However, all those runs are in regime~I; see \Tab{Tsummary_mu}.
We have therefore added the runs of series~X, which are in regime~II.
Nevertheless, the basic slopes are unchanged.  

In \Fig{pmu}(c), we have plotted $\EEGW^{\rm sat}$ versus $\EEM^{\max}/k_\mu$.
This allows us to estimate an upper bound for the empirical parameter $q$
in \Eq{qdef} if $k_{\rm peak}$ is replaced by $k_\mu$.
We find $q<10$.

In view of \Eq{qdef}, using the dependence of $k_{\rm peak}$ on
$v_\mu$ and $v_\lambda$, as given below \Eq{qdef2},
we have $q\propto(\vmu\vlam)^{1/2}$ in regime~II and
$q\propto(\vmu^3/\vlam)^{1/2}$ in regime~I.
This has also been verified using our numerical data.

\subsection{Combined dependence}

\EEq{vlam5vmu} has the advantage that one can now summarize all of the
numerical data in one plot.
The result is shown in \Fig{pall}.
In its inset, we also show $\EEGW^{\rm sat}$ for the set of parameters given by
\cite{BSRKBFRK17} for the early universe, $\vmu=2\times10^{-5}$ and
$\vlam=0.05$, corresponding to $\vlam^5\vmu=6\times10^{-12}$.
It should be noted, however, that those values are rather uncertain,
because both are proportional to $\muz$, for which only 
uncertain upper bounds can be proposed.

Looking at \Fig{pall}, we see that a few runs fall outside the linear
trend.
This applies especially to the runs of series~F and G (red dotted
and blue dotted lines, respectively).
Also the runs of series~U and X (black dashed
and orange dashed lines, respectively) show major departures.
However, it is not immediately obvious what is special about them.

Looking at \Fig{pall}, we see that data points from one series
are identical with data points from another.
This is because those data points are from the same runs, but have alternative names,
see the indications in Tables~\ref{Tsummary_lam} and \ref{Tsummary_mu}.

\subsection{Numerical limitations}

Because of certain numerical constraints,
the parameters of our simulations have 
to stay within specific empirical limits.
The purpose of this section is to discuss the nature of 
those constraints
and to see how they depend on the choice of the parameters.
Let us begin with $\eta$, which we were able to vary by more than four
orders of magnitude.
For smaller values of $\eta$, we go deeper into regime~I, provided
$\eta\lambda^{1/2}<1$.
The main limitation here is the large separation of dynamical and
diffusive time scales.
These time scales are proportional to $\muz^{-1}$ and
$(\eta\muz^2)^{-1}$, respectively.
This separation of time scales results in long run times that make the
simulations more computationally costly.
In addition, there is a large separation in spatial scales between
$\muz^{-1}$ and $\eta$, which corresponds to large magnetic
Reynolds numbers, requiring a large number of mesh points.
And, as we have now seen, for decreasing $\eta$,
the magnetic and GW energies become very small.
For larger $\eta$, on the other hand, we go deeper into regime~II,
provided $\eta\lambda^{1/2}>1$.
The main limitation here is the shortness of the numerical time step,
which depends on the mesh spacing $\delta$ as $\sim\delta^2/\eta$.

Next, let us discuss the value of $\muz$, which we have been able
to vary by a little over two orders of magnitude.
Clearly, for the dynamo instability to exist, the mesh spacing cannot be
too coarse, and $\muz$ must not exceed the largest resolved wavenumber
in the domain $\pi/\delta=k_1 N/2$.
Therefore, for a given number of mesh points $N$, $k_1$ cannot be too small.
It cannot be too large either, because then we would no longer be able
to capture the largest length scales in the system.
In particular, if $k_1$ is too large, it could lead to artifacts
resulting from the finiteness of the domain, as already discussed
in \Sec{LateTimes}.
As we see from \Tab{Tsummary}, we have varied $k_1$ by a factor of 20.
It should be noted that it is not a physical parameter, since the
intention is to simulate an infinitely extended domain.
Therefore, the final results should be independent of $k_1$.
An example is seen by inspecting  \Fig{EEGW_vs_EEKM} for the runs
of series~A, where the three uppermost  open black symbols show a small
shift to the left.
This is because here $k_1$ has been decreased from 100 to 50.
In \Fig{pspec_sat_512_1e2_1e4_4e8_1em6aD}, for example, $k_1$ is not
small enough to capture the maximum GW energy properly.

Finally, the parameter $\lambda$ determines the limiting CME speed
$\vlam$.
We have varied $\lambda^{1/2}$ by over two orders of magnitude.
For the smallest values in \Tab{Tsummary}, we also needed to decrease
the value of $\muz$ to prevent the magnetic energy from exceeding the
critical density, which corresponds to a value of unity.
This could lead to the production of shocks which, in turn, requires
more mesh points, larger viscosity, or both.
Furthermore, the neglect of special relativistic effects could no longer
be justified.

\section{Conclusions}

The present work has revealed a scaling relation for the GW energy
from the CME: $\EEGW^{\rm sat}\propto\vlam^5\vmu$.
Based on earlier dimensional arguments and numerical findings for
the resulting magnetic field energy \citep{BSRKBFRK17}, it was already
anticipated that, within the framework of the standard description of
the CME including its dependence on temperature and the effective number
of degrees of freedom, the resulting GWs would be too weak to
be detectable.
This is indeed confirmed by our present work.
Furthermore, we have also shown that the conversion from magnetic to GW
energy is generally less efficient than for forced and decaying
turbulence; see \Fig{EEGW_vs_EEKM}.
Here, we have been able to estimate the efficiency parameter
$q$ in \Eq{qdef} as being roughly $\propto (\vmu\vlam)^{1/2}$ in regime~II, 
but $\propto (\vmu^3/\vlam)^{1/2}$ in regime~I.
It should also be emphasized that, even though $q$ can reach values
of the order of ten (see Tables~\ref{Tsummary}, \ref{Tsummary_lam},
and \ref{Tsummary_mu}), which is similar to the value for acoustic
turbulence, the final GW energy production is still poor owing to the
small length scales associated with the CME.

Magnetic field generation by the CME can occur in two different regimes;
regimes~I and II, depending on the relation of magnetic field generation
and limiting CME speeds, $\vmu$ and $\vlam$, respectively.
In the present work, we have regarded the CME as a generic mechanism
that allows us to study how GW energy production can be related to the
strengths of generation and the limiting CME speed.
Whether or not other magnetogenesis mechanisms can really be described
in similar ways needs to be seen.
It is interesting to note, however, that our finding regarding the
proportionality of the GW energy to the fifth power of $\vlam$
is reminiscent of the earlier results of \cite{Gogo07} who found the GW
energy to be proportional to the fifth power of the turbulent velocity;
see their Equation~(40).
It should be noted, however,
that the additional dependence on $\vmu$ cannot be neglected and
results in the increase of $\EEGW$ with increasing values of $\eta$;
see \Fig{peta}(b).

Our work has also revealed new unexpected GW energy spectra.
In regime~I, the spectra were not of clean power law form, and the
spectral energy was falling off with wavenumber faster than in any
earlier simulations.
This means that the GW energy $\EEGW=\int \EGW(k)\,\dd k$ depends
significantly on its lower integration bound $k_1$ so that it will be
important to include even smaller wavenumbers in future simulations.
This could restore a quadratic scaling for Runs~A1--A4 and Runs~B1--B5
in Figure~1 and Figure~10(c).
In regime~II, on the other hand, we have seen that large GW energies
can be generated.
This was rather surprising and counterintuitive, because this regime
implies a lack of a turbulent cascade in $\EM(k)$ with just a spectral
bump traveling toward lower wavenumbers.
This traveling, on the other hand, happened rather rapidly, which
contributed to the large GW energies in that case.
The physical reality of this regime is however questionable.

\vspace{12mm}\noindent
{\em Software and Data Availability.} The source code used for
the simulations of this study, the {\sc Pencil Code} \citep{PC},
is freely available on \url{https://github.com/pencil-code/}.
The DOI of the code is https://doi.org/10.5281/zenodo.2315093 \citep{v2018.12.16}.
The simulation setup and the corresponding data are freely available from
\url{https://doi.org/10.5281/zenodo.4448211}; see also
\url{http://www.nordita.org/~brandenb/projects/GWfromCME/} for easier access.

\acknowledgments
Support through grants from the Swedish Research Council (2019-04234),
the Shota Rustaveli National Science Foundation of Georgia (FR18-1462),
and the European Research Council (694896) are gratefully acknowledged.
We acknowledge the allocation of computing resources provided by the
Swedish National Allocations Committee at the Center for Parallel
Computers at the Royal Institute of Technology in Stockholm.
J.S.\ acknowledges the funding from
the Swiss National Science Foundation under Grant No.\ 185863.
The computations and data handling were enabled by resources provided by
the Swedish National Infrastructure for Computing (SNIC) at the Center
for Parallel Computers at the Royal Institute of Technology in Stockholm,
partially funded by the Swedish Research Council through grant agreement
no.\ 2018-05973.

\appendix
\section{The compression term in Equation (2)}
\label{ExtraTerm}

\begin{figure}\begin{center}
\vspace{3mm}
\includegraphics[width=.5\textwidth]{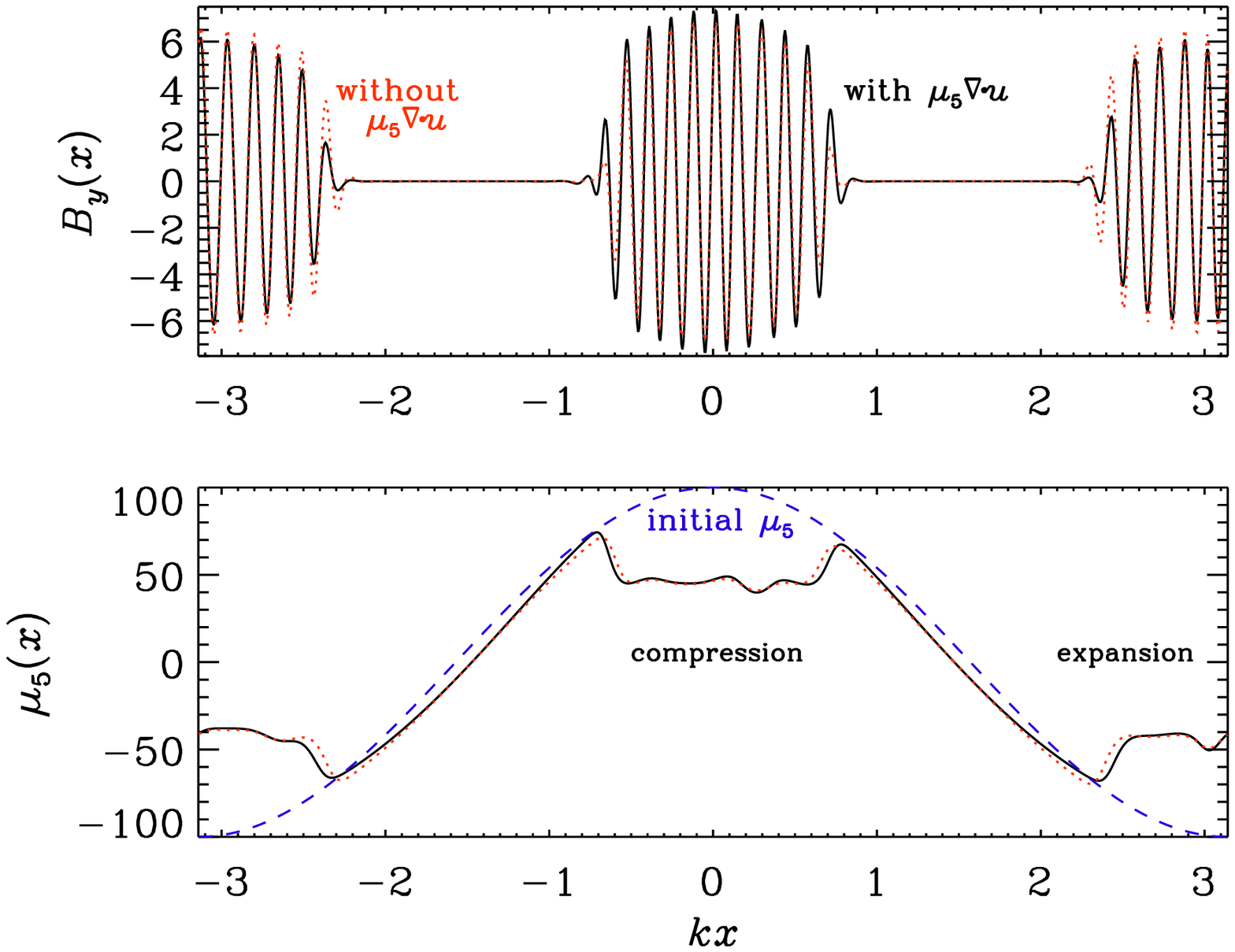}
\end{center}\caption{
Comparison of the profiles of $B_y(x)$ and $\muf(x)$ for $k=1$,
$\muz=100$, $\lambda=100$, $\eta=10^{-3}$, and $u_0=10^{-2}$ with (black)
and without (red) the $\muf\nab\cdot\uu$ term included.
The initial profile of $\muf$ is also shown (blue dashed).
}\label{pcomp_var}\end{figure}

At the end of \Sec{BasicEquations}, we noted that for $\nab\cdot\uu\neq0$,
the conservation of the total chirality requires an extra term,
$-\muf\nab\cdot\uu$, on the right-hand side of \Eq{dmudt}.
For $\Gamma_{\rm\!f}=0$, this equation can then also be written as
\begin{equation}
{\partial\muf\over\partial t}=-\nab\cdot(\muf\uu)
-\lambda\,\eta\left(\muf\BB-\JJ\right)\cdot\BB+D_5\nabla^2\muf,
\label{dmudt2}
\end{equation}
expressing the conservation of $\muf$ for $\BB=\boldsymbol{0}$.
To illustrate the effect of the $\muf\nab\cdot\uu$ term, we consider
here a simple one-dimensional example with a prescribed (kinematic)
velocity field $\uu=(u_0\sin kx, 0, 0)$ and periodic boundary conditions.
This is obviously an artificial way of demonstrating the consequences
for the generation of $\BB$.
To have an effect on the conservation of $\muf$, we also consider an
initial profile of the form $\muf(x,0)=\muz\cos kx$, so
$\bra{\muf(x,0)} =0$.
In \Fig{pcomp_var}, we show $B_y(x)$ and $\muf(x)$ at $t=10$ for $k=1$,
$\muz=100$, $\lambda=100$, $\eta=10^{-3}$, and $u_0=10^{-2}$.
We used a weak seed magnetic field with zero helicity as the initial
condition.

\begin{figure}\begin{center}
\includegraphics[width=.5\textwidth]{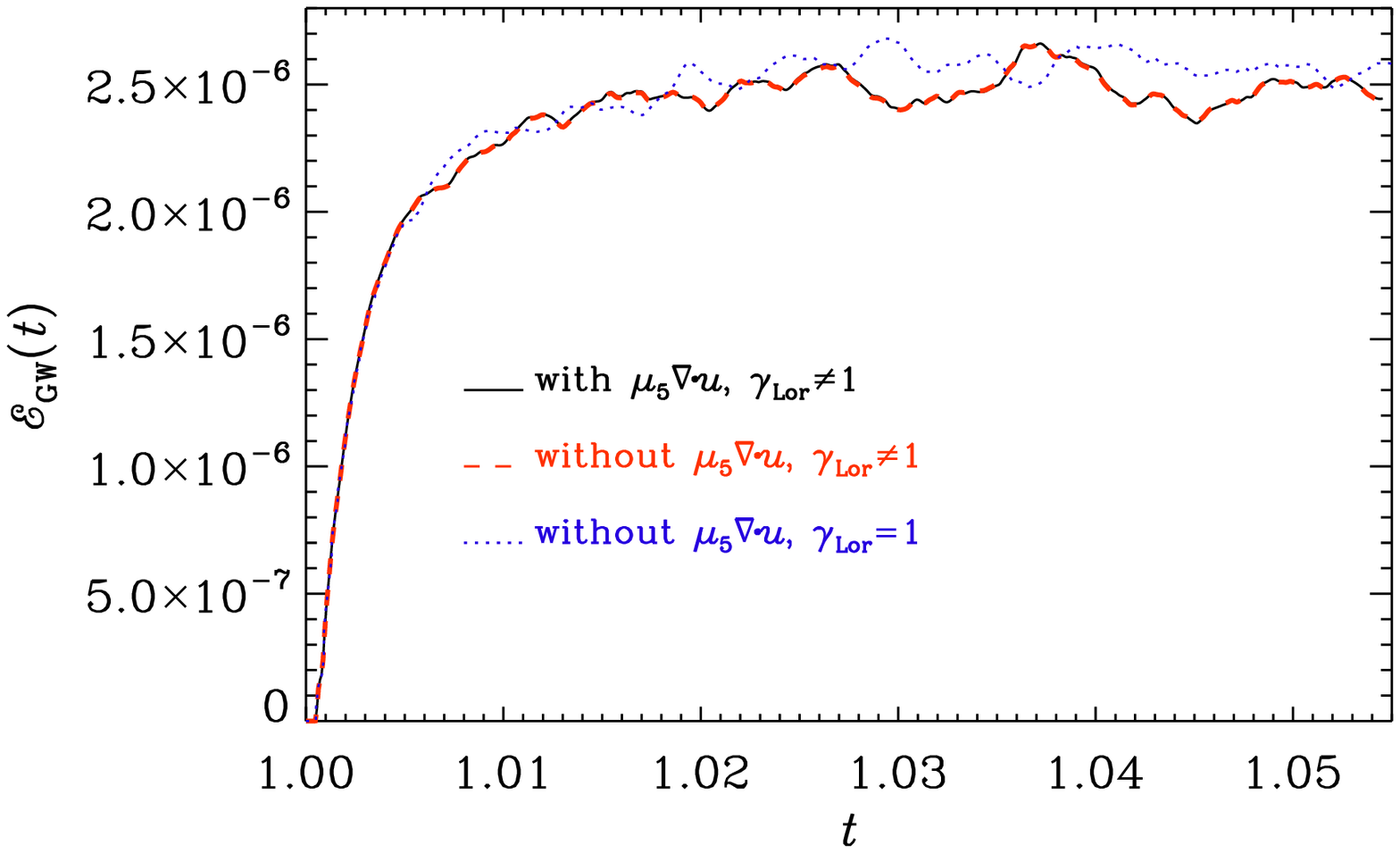}
\end{center}\caption{
Comparison of $\EEGW(t)$ for the cases where $\gamma_{\rm Lor}\neq1$ and
the $\muf\nab\cdot\uu$ term is included (black solid) and where it is
omitted (red dashed) with a case where $\gamma_{\rm Lor}=1$ and the
$\muf\nab\cdot\uu$ term is included (blue dotted).
}\label{pcomp_512_1e2_1e4_2p5e7_2em3a_gam_divu}\end{figure}

When conservation of the total chirality is invoked
by including the $\muf\nab\cdot\uu$ term, there is a small enhancement
of $B_y$ around $kx=\pm0.7$ and a small decrease at $\pm2.3$.
This is caused by compression at $kx=0$ and expansion at $kx=\pm\pi$.
In this example, when the $\muf\nab\cdot\uu$ term is absent, the total chirality 
becomes negative and reaches about 6\% of its initial rms value.   
Finally, we show in \Fig{pcomp_512_1e2_1e4_2p5e7_2em3a_gam_divu} the
evolution of $\EEGW(t)$ for Run~D8, where the magnetic field is one of
the largest and the effect of the $\muf\nab\cdot\uu$ term is expected
to be strong.
We compare the case where $\gamma_{\rm Lor}\neq1$ and the $\muf\nab\cdot\uu$ term
is included with a case where it is omitted, and a case where $\gamma_{\rm Lor}=1$
and the $\muf\nab\cdot\uu$ term is included.
The effect of the latter is here extremely small.
We also see that the inclusion of the $\gamma_{\rm Lor}$ term affects the detailed
time evolution of $\EEGW(t)$, but not the final overall saturation level.

\vspace{3cm}
%r e f


\begin{thebibliography}{}

\bibitem[Aharonian et al.(2006)]{Aharonian}
Aharonian, F., Akhperjanian, A. G., Bazer-Bachi, A. R., et al.\ynat{2006}{440}{1018}

\bibitem[Anand et al.(2019)]{ABPK19}
Anand, S., Bhatt, J. R., Pandey, A. K., \& Kumar, A.\yjour{2019}{Eur. Phys. J. C}{79}{119}

\bibitem[Arnold et al.(2000)]{Arnold00}
Arnold, P., Moore, G. D., \& Yaffe, L. G.\yjhep{2000}{11}{001}

\bibitem[Boyarsky et al.(2021)]{BCRS21}
Boyarsky, A., Cheianov, V., Ruchayskiy, O., \& Sobol, O.\yprd{2021}{103}{013003}

\bibitem[Boyarsky et al.(2012)]{BFR12}
Boyarsky, A., Fr\"ohlich, J., \& Ruchayskiy, O.\yprl{2012}{108}{031301}

\bibitem[Boyarsky et al.(2015)]{BFR15}
Boyarsky, A., Fr\"ohlich, J., \& Ruchayskiy, O.\yprd{2015}{92}{043004}

\bibitem[Brandenburg(2018)]{v2018.12.16}
Brandenburg, A., on behalf of the Pencil Code Collaboration, 2018,
{\sc Pencil Code}, v2018.12.16, Zenodo, DOI:10.5281/zenodo.2315093

\bibitem[Brandenburg \& Boldyrev(2020)]{BB20}
Brandenburg, A., \& Boldyrev, S.\yapj{2020}{892}{80}

\bibitem[Pencil Code Collaboration(2021)]{PC}
Pencil Code Collaboration: Brandenburg, A., Johansen, A., Bourdin, P. A., Dobler, W., Lyra, W., Rheinhardt, M., Bingert, S., Haugen, N. E. L., Mee, A., Gent, F., Babkovskaia, N., Yang, C.-C., Heinemann, T., Dintrans, B., Mitra, D., Candelaresi, S., Warnecke, J., K\"apyl\"a, P. J., Schreiber, A., Chatterjee, P., K\"apyl\"a, M. J., Li, X.-Y., Kr\"uger, J., Aarnes, J. R., Sarson, G. R., Oishi, J. S., Schober, J., Plasson, R., Sandin, C., Karchniwy, E., Rodrigues, L. F. S., Hubbard, A., Guerrero, G., Snodin, A., Losada, I. R., Pekkil\"a, J., \& Qian, C.\yjour{2021}{Journal of Open Source Software}{6}{2807}
arXiv:2009.08231, doi: 10.21105/joss.02807

\bibitem[Brandenburg \& Kahniashvili(2017)]{BK17}
Brandenburg, A., \& Kahniashvili, T.\yprl{2017}{118}{055102}

\bibitem[Brandenburg et al.(1996)]{BEO96}
Brandenburg, A., Enqvist, K., \& Olesen, P.\yprd{1996}{54}{1291}

\bibitem[Brandenburg et al.(2017a)]{BKMRPTV17}
Brandenburg, A., Kahniashvili, T., Mandal, S., Roper Pol, A., Tevzadze, A. G., \& Vachaspati, T.\yprd{2017a}{96}{123528}

\bibitem[Brandenburg et al.(2017b)]{BSRKBFRK17}
Brandenburg, A., Schober, J., Rogachevskii, I., Kahniashvili, T., Boyarsky, A., Fr\"ohlich, J., Ruchayskiy, O., \& Kleeorin, N.\yapjl{2017b}{845}{L21}

\bibitem[Caprini et al.(2019)]{Caprini19}
Caprini, C., Chala, M., Dorsch, G. C., Hindmarsh, M., Huber, S. J., Konstandin, T., Kozaczuk, J., Nardini, G., No J. M., \& Rummukainen, K.\yjcap{2019}{03}{024}

\bibitem[Deryagin et al.(1987)]{Deryagin}
Deryagin, D. V., Grigoriev, D. Y., Rubakov, V. A., \& Sazhin, M. V.\ymn{1987}{229}{357}

\bibitem[D{\'{\i}}az-Gil(2008a)]{DiasGil08a}
D{\'{\i}}az-Gil, A., Garc{\'{\i}}a-Bellido, J., Garc{\'{\i}}a P{\'e}rez, M., \& Gonz{\'a}lez-Arroyo, A.\yprl{2008a}{100}{241301}

\bibitem[D{\'{\i}}az-Gil(2008b)]{DiasGil08b}
D{\'{\i}}az-Gil, A., Garc{\'{\i}}a-Bellido, J., Garc{\'{\i}}a P{\'e}rez, M., \& Gonz{\'a}lez-Arroyo, A.\yjour{2008b}{J. High Energy Phys.}{2008}{07}{043}

\bibitem[Durrer \& Caprini(2003)]{DC03}
Durrer, R., \& Caprini, C.\yjour{2003}{JCAP}{0311}{010}

\bibitem[Durrer \& Neronov(2013)]{DN13}
Durrer, R., \& Neronov, A.\yanar{2013}{21}{62}

\bibitem[Gogoberidze et al.(2007)]{Gogo07}
Gogoberidze, G., Kahniashvili, T., \& Kosowsky, A.\yprd{2007}{76}{083002}

\bibitem[Grasso \& Rubinstein(2001)]{GR01}
Grasso, D., \& Rubinstein, H. R.\ypr{2001}{348}{163}

\bibitem[Hindmarsh et al.(2015)]{HHRW15}
Hindmarsh, M., Huber, S. J., Rummukainen, K., \& Weir, D. J.\yprd{2015}{92}{123009}

\bibitem[Joyce \& Shaposhnikov(1997)]{JS97}
Joyce, M., \& Shaposhnikov, M.\yprl{1997}{79}{1193}

\bibitem[Kahniashvili et al.(2021)]{Kahn21}
Kahniashvili, T., Brandenburg, A., Gogoberidze, G., Mandal, S., \& Roper~Pol, A.\pprr{2020}
{2011.05556}

\bibitem[Kahniashvili et al.(2013)]{KTBN13}
Kahniashvili, T., Tevzadze, A. G., Brandenburg, A., \& Neronov, A.\yprd{2013}{87}{083007}

\bibitem[Kosowsky et al.(2002)]{KMK02}
Kosowsky, A., Mack, A., \& Kahniashvili, T.\yprd{2002}{66}{024030}

\bibitem[Okano \& Fujita(2021)]{OF21}
Okano, S., \& Fujita, T.\yjcap{2021}{03}{026}

\bibitem[Neronov \& Vovk(2010)]{NV10}
Neronov, A., \& Vovk, I.\ysci{2010}{328}{73}

\bibitem[Neronov et al.(2021)]{Neronov20}
Neronov, A., Roper Pol, A., Caprini, C., \& Semikoz, D.\yprd{2021}{103}{L041302}

\bibitem[Rogachevskii et al.(2017)]{Roga17}
Rogachevskii, I., Ruchayskiy, O., Boyarsky, A., Fr\"{o}hlich, J., Kleeorin, N., Brandenburg, A., \& Schober, J.\yapj{2017}{846}{153}

\bibitem[Roper Pol et al.(2020a)]{RPBKKM20}
Roper Pol, A., Brandenburg, A., Kahniashvili, T., Kosowsky, A., \& Mandal, S.\ygafd{2020a}{114}{130}

\bibitem[Roper Pol et al.(2020b)]{RPMBKK20}
Roper Pol, A., Mandal, S., Brandenburg, A., Kahniashvili, T., \& Kosowsky, A.\yprd{2020b}{102}{083512}

\bibitem[Sharma et al.(2020)]{SSS19}
Sharma, R., Subramanian, K., \& Seshadri, T. R.\yprd{2020}{101}{103526}

\bibitem[Schober et al.(2018)]{Schober18}
Schober, J., Rogachevskii, I., Brandenburg, A., Boyarsky, A., Fr\"{o}hlich, J., Ruchayskiy, O., \& Kleeorin, N.\yapj{2018}{858}{124}

\bibitem[Schober et al.(2020)]{Schober20}
Schober, J., Brandenburg, A., \& Rogachevskii, I.\ygafd{2020}{114}{106}

\bibitem[Taylor et al.(2011)]{TVN11}
Taylor, A. M., Vovk, I., \& Neronov, A.\yana{2011}{529}{A144}

\bibitem[Vilenkin(1980)]{Vil80}
Vilenkin, A.\yprd{1980}{22}{3080}

\end{thebibliography}
\end{document}